\documentstyle[11pt,aaspp4]{article} 

\def\chaphead{}

\def\spose#1{\hbox to 0pt{#1\hss}}
     
\def\={\overline}

\newcount\notenumber
\newcount\eqnumber
\newcount\fignumber
\newbox\abstr
\newbox\figca

\def\etal{{\it et al. }}
\def\eg{{\it e.g., }}
\def\ie{{\it i.e., }}
\def\cf{{\it cf. }}

\def\note#1{\footnote{$^{\the\notenumber}$}{#1}\global\advance\notenumber by 1}
\def\foot#1{\raise3pt\hbox{\eightrm \the\notenumber}
     \hfil\par\vskip3pt\hrule\vskip6pt
     \noindent\raise3pt\hbox{\eightrm \the\notenumber}
     #1\par\vskip6pt\hrule\vskip3pt\noindent\global\advance\notenumber by 1}

\def\Dt{\spose{\raise 1.5ex\hbox{\hskip3pt$\mathchar"201$}}}    
\def\dt{\spose{\raise 1.0ex\hbox{\hskip2pt$\mathchar"201$}}}    

\def\new{{\rm\chaphead\the\eqnumber}\global\advance\eqnumber by 1}
\def\ref#1{\advance\eqnumber by -#1 \chaphead\the\eqnumber
     \advance\eqnumber by #1 }
\def\last{\advance\eqnumber by -1 {\rm\chaphead\the\eqnumber}\advance
     \eqnumber by 1}
\def\eqnam#1{\xdef#1{\chaphead\the\eqnumber}}
     
\def\nfig{\chaphead\the\fignumber\global\advance\fignumber by 1}
\def\nfiga#1{\chaphead\the\fignumber{#1}\global\advance\fignumber by 1}
\def\rfig#1{\advance\fignumber by -#1 \chaphead\the\fignumber
     \advance\fignumber by #1}
\def\refindent{\par\noindent\parskip=4pt\hangindent=3pc\hangafter=1 }

\def\apj#1#2#3{\refindent#1,  {ApJ,\ }{#2}, #3}
\def\apjsup#1#2#3{\refindent#1,  {ApJS\ }{#2}, #3}

\def\apjlett#1#2#3{\refindent#1,  { ApJL,\  }{#2}, #3}

\def\mnras#1#2#3{\refindent#1,  { M.N.R.A.S., }{#2}, #3}
\def\annrev#1#2#3{\refindent#1, { ARA \& A,\ }
{\bf2}, #3}
\def\aj#1#2#3{\refindent#1,  { AJ,\  }{#2}, #3}

\def\aa#1#2#3{\refindent#1,  { AA,\ }{#2}, #3}
\def\nature#1#2#3{\refindent#1,  { Nature,\ }{#2}, #3}

\def\pasp#1#2#3{\refindent#1,  { PASP,\ }{#2}, #3}

\def\refbook#1{\refindent#1}

\def\ltsim{\mathrel{\spose{\lower 3pt\hbox{$\mathchar"218$}}
     \raise 2.0pt\hbox{$\mathchar"13C$}}}
\def\gtsim{\mathrel{\spose{\lower 3pt\hbox{$\mathchar"218$}}
     \raise 2.0pt\hbox{$\mathchar"13E$}}}
     
\def\apequal{\mathrel{\spose{\lower 1pt\hbox{$\mathchar"218$}}
     \raise 2.0pt\hbox{$\mathchar"218$}}}
\newbox\grsign \setbox\grsign=\hbox{$>$} \newdimen\grdimen \grdimen=\ht\grsign
\newbox\simlessbox \newbox\simgreatbox
\setbox\simgreatbox=\hbox{\raise.5ex\hbox{$>$}\llap
     {\lower.5ex\hbox{$\sim$}}}\ht1=\grdimen\dp1=0pt
\setbox\simlessbox=\hbox{\raise.5ex\hbox{$<$}\llap
     {\lower.5ex\hbox{$\sim$}}}\ht2=\grdimen\dp2=0pt

\def\eg{{\it e.g.,\ }}
\def\ie{{\it i.e.,\ }}
\def\etal{{\it et al.\ }}
\begin{document} 
\title{\bf Constraints on Intervening Stellar Populations Toward
the Large Magellanic Cloud}

\author{Dennis Zaritsky}
\affil{UCO/Lick Observatories and Department of Astronomy and
Astrophysics,} 
\affil{Univ. of California at Santa Cruz, Santa Cruz, CA, 95064}
\affil{Electronic mail: dennis@ucolick.org}
\author{Stephen A. Shectman and Ian Thompson}
\affil{Carnegie Observatories, 813 Santa Barbara St., Pasadena, CA
91101}
\affil{Electronic mail: shec@ociw.edu,ian@ociw.edu}
\author{Jason Harris and D.N.C. Lin}
\affil{Department of Astronomy and Astrophysics, Univ. of California
at Santa Cruz,}
\affil{Santa Cruz, CA, 95064}
\affil{Electronic mail: jharris@ucolick.org,lin@ucolick.org}

\abstract{The suggestion by Zaritsky \& Lin (1997; ZL) that a vertical
extension of the red clump feature in color-magnitude diagrams (CMDs) of the
Large Magellanic Cloud (LMC) is consistent with
a significant population of foreground stars to the LMC that
could account for the observed microlensing optical
depth (Renault \etal 1997;
Alcock \etal 1997a) has been challenged by various investigators (cf.
Alcock \etal 1997b, Gallart 1998, Bennett 1998, Gould 1998, Beaulieu \&
Sackett 1998, and Ibata, Lewis, \& Beaulieu 1998). We respond by
(1) examining each of the challenges presented to
determine whether any or all of those arguments invalidate the claims
made by ZL and (2) presenting new 
photometric and spectroscopic data obtained in an attempt to resolve this
issue. We systematically discuss why the objections raised so far do not
unequivocally refute ZL's claim. We conclude that
although the CMD data do not mandate the existence of a foreground
population, they are entirely consistent with a foreground population
associated with the LMC
that contributes significantly ($\sim$ 50\%) to the observed
microlensing optical depth.
From our new data, we conclude that $\ltsim$ 40\% of the VRC stars are
young, massive red clump stars because (1) 
synthetic color-magnitude diagrams created using the star 
formation history derived indepdently from HST data (Geha \etal 1998)
suggest that $<$ 50\% of the VRC stars are young, massive red clump
stars,
(2) the angular distribution of the VRC stars is more 
uniform than that of the young (age $<$ 1 Gyr) main sequence stars,
and (3) the velocity dispersion of the VRC stars in the
region of the LMC examined by ZL, $18.4\pm 2.8$ km 
sec$^{-1}$ (95\% confidence limits), is inconsistent with the 
expectation for a young disk population. Each of these arguments 
is predicated on assumptions and the conclusions are uncertain.
Therefore, an exact determination of the contribution to 
the microlensing optical depth by the various hypothesized
foreground populations, and the subsequent conclusions regarding
the existence of halo MACHOs, requires a detailed knowledge of many
complex astrophysical issues, such as the IMF, star formation history,
and post-main sequence stellar evolution.}

\bigskip

\noindent

\section{Introduction}

The solution of the dark matter problem is a fundamental
goal of current astronomical research. A particularly novel and
ambitious approach involves the detection of gravitational lensing
due to dark objects in the Galactic halo passing between us and 
background sources such as the stars in the Magellanic Clouds 
(Paczy\'nski 1986).
Several groups (EROS, cf. Aubourg \etal 1993; MACHO, cf. Alcock \etal
1997a; and OGLE, cf. Udalski, Kubiak, Szymanski 1997) 
have undertaken multi-year observing programs
to detect such lensing
events toward the Small and Large Magellanic Clouds (SMC and LMC). 
The fundamental
successes of those surveys are that they convincingly identify
microlensing events and that their derived microlensing
optical depths are consistent. Using
standard models  for the stellar populations of the Galaxy and LMC,
the observed microlensing 
optical depth, $\tau_\mu$, implies that $\sim$ 50\% of the Galactic 
halo dark matter out to the radius of the LMC is in the form of
massive compact 
objects, MACHOs (Alcock \etal 1997a). 
The importance of this conclusion to our understanding
of a wide range of astrophysical topics is manifest. Is this
conclusion robust?

Several authors have by now questioned various assumptions
leading to this conclusion. 
For example, Sahu (1994) proposed that LMC self-lensing (lensing
of LMC stars by other LMC stars)
is greater than in the standard model and can be sufficient to account for 
$\tau_\mu$;
Zhao (1998a) proposed that an associated intervening 
stellar population, such as a dwarf galaxy along the line of sight, 
might provide a significant fraction of the 
total optical depth; and Evans \etal (1998) proposed that the Galactic
disk may be sufficiently warped and flared to provide $\sim$ 50\%
of $\tau_\mu$. 
Zaritsky \& Lin (1997; hereafter
ZL) proposed that the vertical extension of the red clump feature (termed the
VRC) observed in a million star color-magnitude diagram, CMD, of an 
LMC region observed as part of the Magellanic Clouds Photometric
Survey (MCPS, Zaritsky, Harris \& Thompson 1997; 
hereafter ZHT) is consistent with 
the existence of either an
associated or unassociated 
intervening stellar population that is as much as 15 kpc
closer to us than the LMC and that this population has sufficient
mass density to account for a large fraction, and possibly all, 
of $\tau_\mu$. 

To reach this conclusion, ZL
subtracted a simple model of the red clump distribution and found
a residual population that, if foreground, corresponds to stars
$\sim 15$ kpc closer to us than the LMC. This model greatly
simplified the analysis of the possible effect of an intervening
population, but has led to some confusion in the literature. Because
the VRC is a continuous extension of the red clump toward brighter
magnitudes, it may arise from stars distributed continuously
between 0 and 15 kpc from the LMC. Therefore, the interpretation
of the VRC impacts a variety of intervening-population
scenarios,
ranging from that of a detached 
stellar population unassociated with the LMC to that of a gravitationally
bound halo or ``thick'' disk LMC population. 
The theoretical claims and possible
observational evidence for intervening populations
triggered a flurry of rebuttal papers 
(Alcock \etal 1997b,
Gallart 1998, Bennett 1998, Gould 1998, Johnston 1998, Beaulieu \&
Sackett 1998, Ibata, Lewis, \& Beaulieu 1998).

We systematically re-examine the possibility 
of a foreground population
in two ways: (1) we test each of the arguments presented 
in the rebuttal papers to 
determine whether any or all of those arguments invalidate the 
most general claims
made by ZL and (2) we present new photometric and spectroscopic data
in an attempt to resolve this issue. We define our interpretation of 
Occam's razor in \S1.1.
In \S 2 we describe our new
data, in \S3 we discuss the rebuttal papers and examine their 
conclusions, and in \S4 we discuss the implications of the new
data presented in \S2. The discussion in \S3 illustrates how difficult
it is to eliminate the possibility of a foreground
population responsible for the microlensing. In \S4 we
demonstrate that the nature of the VRC is complex and that stars
in that region of the CMD must have a variety of origins. The myriad
of uncertainties
in the interpretation of the data and in the models applied to 
convert the observations into a microlensing optical depth 
must propagate into an uncertainty in any interpretation of the
microlensing events. We conclude that due to the unresolved
systematic uncertainties 
the microlensing data are yet unable to demonstrate the existence of 
halo MACHOs.

\subsection{Occam's Razor Revisited} 

Because we find 
microlensing by normal stars distributed in unexpected ways to be
more plausible than microlensing by unknown objects distributed
in a smooth halo, we assert that a convincing case for halo MACHOs
can only be made after the sum of all plausible stellar populations is
categorically eliminated as the source of $\tau_\mu$.
Therefore, 
our approach is to aggressively attempt to reconcile the possibility
of a foreground population that accounts for a significant fraction ($\sim
50$\%) of $\tau_\mu$ with the various challenges 
raised by the rebuttal papers. If we can accommodate such
a foreground population within parameter ranges allowed by observational
or model uncertainties, then the interpretation of the microlensing
optical depth is dominated by uncertainties in the parameters
of the foreground population. Given the 
impact of the existence of halo MACHOs on a wide variety of 
disciplines, we believe that 
it is insufficient to argue that the MACHO interpretation
is preferable to other interpretations 
that make moderate adjustments to such
ill-determined quantities as the field initial mass function in the LMC
or the star formation history of
the LMC. We consider arguments that present alternative
interpretations for the VRC population or that conclude that the
optical depth from such a population may be negligible ($\ltsim 10$\%), 
but which
do not {\it convincingly eliminate} the 
presence or effect of such intervening
populations on $\tau_\mu$, to be incomplete (even though they may
be correct). We cede from the beginning that models exist for
which the effect of the 
intervening populations are negligible --- we 
address whether such models are unavoidable.

\section{Data}

\subsection{Photometry}

The photometric data come from an ongoing survey of the
Magellanic Clouds (ZHT).
We have now reduced data from
an area in the LMC that is 
more than three times larger than that available to ZL. The 
new area also contains one of the most vigorous sites of recent ($< 1$
Gyr) star formation in the LMC,
the Constellation III area (Shapley 1956), and so provides a range
of stellar populations with different
star formation histories and stellar densities. Photometric incompleteness
becomes a serious factor only for $V > 21$ or in high density regions,
such as the centers of stellar clusters (ZHT).
Incompleteness is not a problem for this study of 
red clump or upper ($V<19.5$) main sequence stars.
The area being discussed, Figure 1, is
centered at approximately $\alpha = 5^h 20^m$ and
$\delta = -66^\circ 48^\prime$ and has an irregular
shape due to idiosyncrasies of the survey that do not affect our
current analysis. The area discussed by ZL is roughly the lower right
quarter of Figure 1. 
The stellar catalog consists of $\alpha, \delta$ (2000.0) and
$UBVI$ photometry and associated uncertainties. Stars must be detected
in $B$ and $V$ to enter the catalog, but not necessarily in $U$ and
$I$. The current catalog contains $\sim$ 2.5 million stars with $V <
21$ and 4 million in total. 

\subsection{Spectroscopy}

The spectroscopic data were obtained using the Las Campanas 2.5m du Pont
telescope and the multifiber spectrograph (Shectman \etal 1992) during the
nights of January 19-23, 1998. This spectrograph obtains up to 
128 spectra simultaneously. Due to the need to compromise
between maximum spectral resolution and signal-to-noise, we used different
observational setups for the brighter VRC stars than for the 
fainter red clump (RC) stars. 
We used a 1200 l/mm grating 
blazed at 5000\AA\ in 
second order to observe the VRC stars (2.3 \AA\ resolution) and in first order
to observe the RC stars (3.8 \AA\ resolution).
The spectrograph uses a 2D-Frutti detector, which is
blue sensitive, so we concentrate on the spectral region between 
3800\AA\ and 4800\AA. Therefore, velocities are measured 
primarily using the Ca H and K lines. 

One of our primary concerns for these observations
is the contamination of spectra by other 
stars within the 3 arcsec fiber aperture. The fields in the LMC are crowded and
some stars, typically fainter than VRC or RC stars, are always 
present within a few arcsec of a random location.
To quantify this problem and
statistically remove ``sky'' from object spectra, we obtained
spectra at a position offset by 8 arcsec from the 
target position. The offset frames were interleaved between
target observations. The exposure times for individual 
exposures were 3000 sec for target
and 1000 sec for sky. Several exposures, both
target and sky, were taken for each field. The
total exposure times  were 11000 sec for VRC Field 1 
($\alpha = 5^h 6^m$ $\delta = -67^\circ 42^\prime$),
20000 sec for VRC Field 2 ($\alpha = 5^h 11^m$ $\delta = -67^\circ
5^\prime$),
 and 10000 for RC field 1 ($\alpha = 5^h 6^m$ $\delta = -67^\circ 42^\prime$).
Each set of target and sky observations are bracketed by calibration
spectra (internal hollow cathode lamp exposures).
Dark frames and fiber throughput calibration frames (incandescent
lamp exposures) were taken before each night's observations. 

The initial selection of targets was done using the
color-magnitude diagrams originally used to identify the VRC
(ZL). The VRC is defined
using $3.1 <C< 3.4$ and $ 18 <V< 18.75$, where $C \equiv
0.565(B-I) + 0.825(U-V+1.15)$. The quantity $C$ was 
defined by ZL to remove a slight color-magnitude dependence of the red clump.
In practice, it is only a minor modification and selecting the VRC in
the $V-I, V$ space would produce a similar sample of stars.
The RC region is defined
by $3.1 <C< 3.4$ and $ 19 <V< 19.3.$ The definitions for the
two populations are more restrictive in color
than those used by ZL and elsewhere in
this paper to minimize contamination by other populations.
We observe two non-overlapping regions on the sky
located in the region of the LMC discussed by ZL.
The stars were ranked in terms
of isolation from other bright stars.
During the otherwise random fiber assignment,
the most isolated stars are given preference to minimize
contamination.

The data reduction consists of the following steps: (1) normalize long
dark frames to corresponding exposure time and subtract from all
frames, (2) trace the incandescent lamp spectra (to define the
apertures on the detector using a high S/N exposure) and extract it
(to measure the relative throughputs of the fibers), (3) extract
object and sky apertures using the apertures as defined by the
incandescent lamp exposures only allowing a recentering of the
reference apertures, (4) correct for fiber throughput
differences by applying the
correction factors measured from the incandescent lamp exposure,
(5) derive wavelength solution for each fiber using calibration
spectra obtained just before and after the science exposures,
(6) apply wavelength solution to sky and object apertures,
(7) combine sky spectra to form a single average sky spectrum for
each sky position, (8) subtract sky spectrum from each object spectrum
(after correcting for relative exposure times of sky and object), and
(9) combine corresponding object spectra from multiple exposures of
the same field taken during a night. The final product is a set of 
one-dimensional spectra of VRC and RC stars.

We measure velocities using standard cross-correlation software
(XCSAO in IRAF). The template spectrum is constructed from the data
themselves. We use a spectrum of a Galactic K giant to measure 
initial velocity estimates 
for the VRC stars. The spectra of VRC stars with reliable 
correlation velocities ($R \ge 6$), are corrected back to zero
radial velocity and combined to create a template with the same
spectral characteristics of the sample stars (thereby avoiding template
mismatch problems). We iterate this procedure until
the number of reliable velocity measurements converges. 
The final template
spectrum is shown in Figure 2. Because of this procedure, the mean velocity
measured for the VRC stars
is 0 km sec$^{-1}$. The RC stars could possibly have a 
mean velocity offset, and Galactic stars should have $v \sim -$275 km sec$^{-1}$
(measurements of the LMC
systematic velocity vary by only a few km sec$^{-1}$: 274 km sec$^{-1}$ from 
H I observations of Luks \& Rohlfs (1992) and 278 km sec$^{-1}$ from a
sample of planetary nebulae observed by Meatheringham \etal 1988).

It is critical to determine the observational uncertainties
precisely 
because both the mean velocity difference between the VRC and RC stars
and the velocity dispersions of the VRC and RC stars are expected to 
be comparable to the velocity uncertainty
in any single spectrum ($\sim 10$ km sec$^{-1}$).
The cross correlation
analysis calculates a measure of the uncertainty based on the strength
of the correlation peak relative to the noise of the correlation
function, but this calculation can underestimate the uncertainties ---
especially
when there are systematic errors (eg. unstable wavelength solutions
and poor sky subtraction). We conduct several tests to
determine the reliability of the calculated uncertainties: (1)
calibration spectra are cross-correlated against other calibration
spectra to test the stability of the wavelength solution, (2)
the velocities obtained for target stars reduced in the standard
manner are compared with the
velocities obtained when no sky spectra are subtracted to determine the
maximum errors caused by poor sky subtraction, and
(3) the velocities of VRC stars in Field 2 that were 
observed on two nights are compared. 

The first test involves cross-correlating the self-calibrated 
calibration lamp spectra. Ideally, these would all have zero relative
velocity, but in practice because of the sparse sampling of the
wavelength solution, line centering errors, and polynomial fitting to
the dispersion function, the spectra can appear to be shifted relative
to one another. We have randomly selected one aperture from one exposure
to act as a velocity template and cross correlated calibration lamps
from various nights and all the apertures with this one template. 
The distribution of velocities has a mean of $-3.4 \pm 0.4$ km sec$^{-1}$
and a dispersion of $2.6 \pm {0.3}$ km sec$^{-1}$. The distribution
is Gaussian (see left panel Figure 3). The uncertainties introduced by errors
in the wavelength calibration, especially because each stellar velocity 
consists of at least four different exposures (each with a different
wavelength calibration) is negligible.

The second test is aimed at determining the uncertainties introduced
by sky subtraction and contamination. We compare the results
from one target field, with the results for the same stars when no 
sky subtraction is performed. This comparison 
should provide an upper limit to the
velocity uncertainty associated with contamination and poor sky 
subtraction. 
As shown in the right panel of Figure 3, the distribution of velocity
differences is non-Gaussian and centered off zero
(as expected, because the sky contributes a variable amplitude, 
but 
constant velocity signal to the spectrum). Because the best fit Gaussian
has a dispersion of 5 km sec$^{-1}$, sky subtraction
is not as serious a problem as was originally envisioned. The
contribution to the overall dispersion, which should be less than 
in this extreme scenario, must be well less than 5 km sec$^{-1}$ added in 
quadrature.

The third test involves comparing the measured velocities
for the stars in the one target
field that was observed on separate nights.
We apply the criteria
that $R \ge 6$ for an acceptable velocity, as we apply for all of
our velocity measurements. 
We evaluate $D \equiv v_1 - v_2$, where
$v_1$ and $v_2$ are the measured velocities from the first
and second night respectively.
The distribution of $D$ should be Gaussian
with mean zero and dispersion given by the standard error propagation,
$\sigma^2 = \sigma_{v1}^2 + \sigma_{v2}^2$. The distribution of 
$D^{\prime} \equiv D/ \sigma$ should be Gaussian with mean zero and 
a dispersion of one if there are no systematic errors and the
uncertainties calculated by the cross-correlation software are accurate. 
For the 
80 stars that satisfy our correlation criteria in both
night's data, the distribution
is well fit by a Gaussian ($\chi^2 = 0.3$) with a mean of 
$0.0\pm 0.6$
(indicating no significant nightly zero point shift) and
a dispersion of $1.2^{+0.6}_{-0.3}$ (Figure 4). 
This result indicates that the
uncertainties generated by the cross-correlation package are 
consistent with the scatter present in the velocities determined
from different nights and that these uncertainties are reliable
estimates of the true uncertainties. The data, including ($\alpha,
\delta$), $V$ and $B$ magnitudes, velocities, and velocity
uncertainties, are
available from the first author.
The conclusions we draw from the data are discussed in \S 4.

\section{Discussion}

\subsection{Previous Studies}

A variety of studies, mostly in response to ZL, 
challenge the suggestion that foreground material traced
by the VRC could account for $\tau_\mu$.
The argument presented by ZL consisted of several steps, 
and each has been investigated in the series of papers discussed
below. In this section, we discuss the studies by
Alcock \etal (1997b), Gallart (1998), Bennett (1998), Gould (1998),
Johnston (1998), and Beaulieu \& Sackett (1998) 
in chronological order of publication. We discuss the results
of Ibata, Lewis, \& Beaulieu (1998) in \S4 in the context of our radial
velocity measurements.

\subsubsection{Alcock \etal 1997b}

The most direct test for the presence of intervening populations
is to measure distances to stars (or lenses) along the line of sight.
Alcock \etal use the MACHO database to search for variable 
stars for which distances can be derived (Cepheids and RR Lyrae). 
They do not find a population of foreground Cepheids, but
Cepheids are somewhat unlikely in an intervening population since
their presence would imply very recent (age $\ltsim {\rm few} \times 10^8$ yrs;
Kippenhahn \& Smith 1969; Efremov 1978; Grebel \& Brandner 1998) 
star formation.
RR Lyrae are more likely to be present in an intervening population and 
Zhao (1998a) noted that the data of Payne-Gaposchkin (1971) contained
a tantalizing clump of RR Lyrae candidates at a distance of $\sim$ 20
kpc. 

Alcock \etal found no excess
of RR Lyrae over the number expected in a flattened ($b/a = 0.6$)
halo model that was normalized to produce the total number of stars
identified to a distance of 40 kpc along the line of sight to the
LMC. However, due to possible confusion with blends, LMC RR Lyrae, and
other variable stars they imposed a selection cut of $V < 18$. For
their adopted RR Lyrae absolute V magnitude of $+0.4$, 
this selection cut corresponds to a limiting distance of
33 kpc. Therefore, while their result argues against
the Zhao hypothesis of an intervening galaxy at $\sim 20$ kpc,
it does not place strong limits on possible stellar populations at 
distances of 35 to 50 kpc. 

Alcock \etal also note that there are no tell-tale features of a foreground
population visible in deep HST CMDs. Because of the expected
large number of low mass stars for 
each ``high'' mass star (\eg red clump star),
a foreground population
may be more statistically distinct along the lower main sequence or
main sequence turnoff region of a CMD than at brighter magnitudes. However,
such comparisons, when the hypothesized foreground 
population is a small fraction of
the total population, are difficult because of the nearly vertical
distribution of main sequence stars in CMDs. In Figure 5 we compare
an original HST CMD of the LMC (courtesy of J. Holtzman from
Holtzman \etal 1997) with a CMD in which
an additional component that consists of 8\% of the original stars
was placed uniformly between 10 and 15 kpc closer than the LMC
along the line of sight. 
The two CMDs are nearly indistinguishable (this degeneracy
in part motivated the suggestion of an intervening
population near the LMC by Zhao (1998a)). The additional freedom 
regarding relative metallicities and reddenings for the foreground
population would enable one to create even less distinguishable 
CMDs (Geha, priv. comm.). Therefore, although deep CMDs may
eventually provide valuable constraints on the line of sight stellar
distribution they are currently not able to confirm or refute 
the presence of stars within 15 kpc of the LMC at the fractional level
proposed by ZL.

\subsubsection{Gallart 1998}

The multi-million star CMDs produced by the MCPS and the microlensing
surveys provide a wealth of information on stellar evolution by 
containing statistically significant populations in rare phases of
stellar evolution. An example of this claim
is an overdensity that was identified
by ZL along the giant branch (in BV CMDs with the observational 
uncertainties of the MCPS, the red giant and asymptotic giant
branches are indistinguishable). ZL associated this particular feature
with the red giant branch bump (RGBB) seen in some globular
clusters (King \etal 1985; Fusi Pecci \etal 1990).  Gallart examined
this feature in detail and identified it as the AGB bump that is
predicted by the Padova isochrones (Bertelli \etal 1994 and
references therein). She further demonstrated that
the RGBB would appear at fainter magnitudes than the red clump
in the LMC population. This revision of the identification of
the feature along the red giant branch is important in terms of
stellar evolution and star formation models of the LMC, but does not
impact the hypothesis of an intervening population --- except as a 
cautionary tale 
of the possible manifestation of previously unidentified
stellar evolutionary phases.

Gallart also noted on the basis of synthetic CMDs that one
should expect a population of stars extending vertically from the
red clump for stellar populations with a component that is younger
than 1 Gyr and that such a feature is observed in the
Fornax and Sextans A dwarfs (Stetson 1997; Dohm-Palmer \etal 1997). 
This suggestion has important ramifications for the
interpretation of the VRC, but we postpone exploring this option
until we discuss the work by Beaulieu and Sackett (1998) which 
examined this possibility quantitatively.

\subsubsection{Bennett 1998}

A key step in determining if a candidate foreground population is
an important source of uncertainty
in the interpretation of $\tau_\mu$ is the evaluation of the
corresponding microlensing optical
depth of such a population, $\tau_{fg}$. 
For example, if the population suggested to be associated with the VRC has
$\tau_{fg} \ll \tau_\mu$, then this result is a robust
argument against the importance of this population 
because it
is independent of the nature of the VRC. The 
origin of the VRC, at 
least with respect to microlensing surveys, becomes academic.

The conversion of a VRC surface density into a corresponding 
$\tau_{fg}$ is uncertain because the VRC 
is only a tracer of the possible foreground population.
To convert the number of VRC stars observed into a surface 
mass density of the associated stellar population requires
assumptions about the completeness of the survey,
the distribution of distances among the stellar populations,
and, most importantly, the mass-to-RC star ratio of the population
(which in turn depends on the initial mass function (IMF)
and the star formation history).

ZL adopt an empirical approach
and assume that the proposed foreground stellar population is the 
same as the LMC population (eg. same M/L). They calculate the mass
density per red clump star for the LMC population using the LMC
rotation curve to derive the mass. 
Concerns have been raised about this approach (see
the discussion of Gould (1998) below). 
Once the mass density per red clump star for the LMC is known, ZL
evaluate the surface mass density of the foreground population by attributing
the same mass density per red clump star to each VRC star (additional geometric
considerations were also applied). Once the surface mass density
of the foreground population 
is evaluated the conversion to an optical depth is noncontroversial.
ZL conclude that within
large uncertainties (at least 50\%) 
the observed lensing could be due to the foreground population, thereby
potentially making this stellar population a critical component of the entire
lensing budget along the line of sight to the LMC. Because of
various ``factor-of-two'' uncertainties, this calculation does not
prove that the proposed
foreground population accounts for the observed
lensing, but instead provides a viable plausibility argument. Given
our philosophy (\cf \S1.1), this result is sufficient for us
to question the existence of halo MACHOs.

Bennett critiqued the calculation of $\tau_{fg}$ and
presented an alternate analytic method for calculating 
the mass density of the foreground
population. For a given
IMF and star formation history, one can calculate the number of RC
stars and the mass density of the entire population. 
Bennett presented such a calculation and concluded that {\it even if}
the VRC were truly a tracer of a foreground population, such
a population would {\it at
most} 
contribute $\tau_{fg} = 0.13 \tau_{\mu}$. This result directly contradicts
the ZL calculation by at least a factor of several, perhaps
a factor of 10, and argues that the VRC, regardless of its origin,
is at most a small factor in the interpretation of the results
from the microlensing surveys. 

Because Bennett's conclusion is critical in assessing the nature of
the lensing sources, we examine the calculation in detail. 
His conclusion 
is robust {\it if} it is independent of assumed parameters, within the
plausible range of such parameters.
For example, if a slight change in the adopted IMF slope results in $\tau_{fg}
\sim \tau_{\mu}$, then this approach is indeterminate and
does not provide proof against the possible importance of a foreground
population traced by the VRC.

Because this theoretical
approach requires one to know
the absolute number of VRC stars, rather than the ratio
between VRC and LMC red clump stars as used by ZL, we need an
accurate census of VRC stars. Bennett adopted 
the number of $\sim$ 70,000 red clump stars in the ZL area (2$^\circ \times
1.5^\circ$) and a relative ratio between VRC and RC stars of
5\% to derive a surface density of 1200 ``foreground'' stars per
sq. degree. These are rough estimates from the ZL histograms and here
we take a more precise census.
There are 89849 RC stars for an RC defined to have
$0.7 \le B-V < 1.15$ and $18.85 < V < 19.5$. The area covered
is 2.93 sq. degrees. The VRC fraction was estimated by ZL to be 
between 5 and 7\%, but that number is sensitive to the
model that is used to subtract the red clump contribution from
the VRC and
to the definition of the color and magnitude 
boundaries for the VRC stars. As can be
seen in ZL's Figure 4, the red clump model subtracts a substantial
fraction of the
potential VRC stars. Nevertheless, using 0.07 for the VRC/RC ratio,
which was ZL's upper limit
(Bennett used 0.05 and Beaulieu and Sackett estimate it to be 0.08),
we obtain a surface density of 2,150 ``foreground'' stars per
sq. degree. If instead, we use the definition of the VRC that
is used elsewhere in this paper ($2.85 < C < 3.57$, $17.8 < V < 18.5$,
where $C = 0.565(B-I)+0.825(U-V+1.15)$, \cf ZL), then we
measure a surface density of 2,471 ``foreground'' stars per
sq. degree (and derive VRC/RC $=$ 0.08).
These surface densities are  80\% to 106\% larger than the number adopted by Bennett
and simply illustrate one of the many ``factor of two''
uncertainties that affect this argument, even before
uncertainties in IMF's and star formation histories are considered.

With the measurement of the number of red clump stars in hand, 
we can
calculate the corresponding total number and mass of stars in this 
population. That mass can then be converted into a microlensing
optical depth. 
Despite some problematic assumptions (such as the characterization
of the LMC population as a single age population), we present 
the equivalent derivation to Bennett's calculation, with the 
exception that we leave the IMF slope as a free parameter. 
From Bennett, the total mass in stars in a given population that is 
characterized
by a single current main sequence turnoff mass given by
$m_{to}M_\odot$ is given by
\begin{equation}
M = A(\int_0^{m_1}m_1^{\Gamma}dm + \int^{m_{to}}_{m_1}m^{\Gamma}dm
+\int_{m_{to}}^{m_2} m^{\Gamma-1}m_{wd}dm),
\end{equation}
where $A$ is an arbitrary normalization constant, $m_1$ is the mass
at which the IMF makes the 
transition from a power-law IMF to a flat IMF (\eg for
$m<m_1, n(m)$ is constant) in solar masses, 
$m_2$ is the upper mass cutoff for the
IMF in solar masses, 
$m_{wd}$ is the mass of objects that have evolved past the 
horizontal branch and is taken to be $m_{wd} = (0.15m +
0.38)M_{\odot}$ (Iben and Renzini 1983), and we have expressed the IMF
power-law slope as $\Gamma$. As defined by Scalo (1986) and adopted
by Massey \etal (1995), $\Gamma = d \log \xi(\log m)/d \log m$, where
$\xi(\log m)$ is the number of stars born per unit logarithmic (base
ten) mass interval per unit area per unit time. For reference,
a Salpeter (1955) mass function has $\Gamma = -1.35$.
Equation (1) can be expressed as
\begin{equation}
M = A({{\Gamma}\over{\Gamma+1}}m_1^{\Gamma+1} + {0.85\over \Gamma+1}
m_{to}^{\Gamma+1} - {0.38\over \Gamma}m_{to}^\Gamma +
{{1.5\Gamma+0.38(\Gamma+1)}\over{\Gamma(\Gamma+1)}}10^{\Gamma}).
\end{equation}
The number of horizontal branch (red clump) stars, $N_{RC}$, is given 
by the mass range of stars that are currently sufficiently old to be
on the horizontal branch. Bennett parameterizes that mass interval
as ranging from $m$ to $m(1+\delta)$ where $\delta = t_{RC}/(3.75t_{ms})$.
For an adopted parameterization of the main sequence lifetime, 
$t_{ms} \propto m^{-3.75}$, and a horizontal branch lifetime,
$t_{RC}$, 

\begin{eqnarray}
N_{RC}& = & A\int_{m_{to}}^{(1+\delta)m_{to}}m^{\Gamma-1}dm \nonumber \\
& = &Am^{\Gamma}_{to}\delta.
\end{eqnarray}
Assuming a constant $t_{RC}$ of $10^8$
yrs (following Bennett and adopted from Castellani, Chieffi, \& Pulone
1991) and $t_{ms} = 1.1\times10^{10}m_{to}^{-3.75}$ yr, then 
\begin{equation}N_{RC} = 0.00242Am^{\Gamma + 3.75}_{to}.
\end{equation}
Taking Equation (1) and dividing
by Equation (4) to obtain the total stellar mass per red clump star
gives ,
\begin{eqnarray}
\nonumber M/N_{RC} &=&
{413{\Gamma}\over{\Gamma+1}}m_1^{\Gamma+1}/m_{to}^{\Gamma+3.75} +
{351\over \Gamma+1}
m_{to}^{-2.75} - {157\over \Gamma}m_{to}^{-3.75} \\
&& +{{620\Gamma+157(\Gamma+1)}\over{\Gamma(\Gamma+1)}}10^{\Gamma}m_{to}^{-\Gamma-3.75}.
\end{eqnarray}
The optical depth is calculated by multiplying the surface mass density in
units
of solar masses per square degree by $3.95\times10^{-14}(1-x)/x$ where
$x$ is the fractional distance of the lensing population 
relative to the source population, for an assumed distance to the 
LMC of 50 kpc. 
Therefore, the total optical depth is given by multiplying the
surface number density of VRC stars, which in the foreground
population
hypothesis are the RC stars of this population, by $M/N_{RC}$ and 
by $3.95\times10^{-14}(1-x)/x$. 

To obtain numerical estimates for the optical depth we need to
adopt numerical values for $m_1$, $m_{to}$, $\Gamma$, and $x$.
Bennett examined two values of $m_1$, 0.3 and 0.6, to represent a Galactic
globular cluster type population and a Galactic disk type population,
respectively. The IMF slope $\Gamma$ was chosen to be $-1.3$, where
$n(m) \propto m^{\Gamma-1}$ for $m_1 < m < 10M_\odot$. Selecting
these values, and $x = 0.7$ corresponding to a distance for the 
foreground population of 35 kpc, the optical depth is $< 4\times
10^{-8}$
for $0.9 < m_{to} < 1.7$ in exact agreement with Bennett's Figure 1, 
and is much less than the observed value of
$2.9^{+1.4}_{-0.9}\times 
10^{-7}$  (Alcock \etal 1997a) or the smaller, less certain, 
measurement of Ansari \etal (1996) of
$8.2 \times 10^{-8}$. Analysis of the combined data 
concludes that $\tau_\mu = 2.1^{+1.3}_{-0.8} \times 10^{-7}$ (Bennett 1998).

We now probe the sensitivity of this result on the adopted parameters.
First, we examine the dependence on $m_1$. As $m_1$ decreases, 
$\tau_{fg}$ increases because there are more lower mass stars that 
act as lenses. 
For example, a choice of $m_1 = 0.1$ leads to an increase in the
optical depth of about 70\%. Examining the reference cited for the
choice of $m_1$ (De Marchi \& Paresce 1997), the proper choice of 
$m_1$ does not appear to be precisely determined.
Depending on the adopted mass-luminosity relationship for low mass
stars, the
mass function flattens somewhere between 0.3 and
0.13$M_\odot$. 

Second, we probe the dependence of the calculated $\tau_{fg}$
on the choice of
$\Gamma$. Highly negative values of $\Gamma$ (steeper IMFs) increase the optical
depth because there are more low mass stars for each detected high
mass star. Bennett adopted the standard value and one that appears
to be consistent with observations of LMC associations ($\Gamma =
-1.3\pm 0.3$; Massey \etal 1995). 
However, the field IMF in the LMC appears to be significantly steeper, with
some estimates going as far as $\Gamma = -4.1 \pm 0.2$ (Massey \etal
1995). 
If the appropriate
slope is that steep, we in fact {\it overestimate} the observed
optical depth by
a factor of 2 (even with $m_1 = 0.3$ and Bennett's lower number for
the VRC surface number density). More moderate values of the IMF slope, 
such as $\Gamma = -2$ (with, for example, $m_1 = 0.2$ and the somewhat higher
number density of VRC stars we measure) 
give optical depth estimates in agreement
with those of ZL ($\sim 0.5 \tau_{\mu}$), 
and therefore suggest that the foreground population
{\it could} be an important contributor to the total optical depth.

Admittedly, we have maneuvered the parameters in favor of high
optical depth and the observational determinations of the field
IMF in the LMC are valid only for high mass stars, but the selected
parameters are well 
within the ranges currently allowed by observations.
Even if $\Gamma = -1.6$ (a 1$\sigma$
excursion from the Massey \etal's result for LMC OB associations)
$m_1 = 0.2$, and $m_{to} =
1M_\odot$, $\tau_{fg} = 1 \times 10^{-7}$, 
which is roughly between a third and
half of the observed optical depth 
(Alcock \etal 1997a; Bennett 1998). A combination of a 
comparable population behind the LMC (cf. Zhao 1998b), a warped 
Galactic disk (Evans \etal 1998), and increase contribution
from the known lensing components
(\eg LMC halo and flat galactic disk; Aubourg \etal 1999)
{\it could} account for the
difference between this optical depth and $\tau_{\mu}$.

For allowed parameter ranges, $\tau_{fg}$ can range from insignificant
to $\sim \tau_\mu$.
We find that the uncertainties in the IMF slope and the stellar
mass at which the IMF may flatten, 
preclude the use of this argument to categorically conclude that
a foreground population that has a contrast of $\sim$ 5\% in red clump
stars cannot account for about half of the observed 
optical depth. We stress that Bennett's adopted parameters and 
low optical depths are plausible, perhaps even preferable, but not unique.

As a final note, it is illuminating to calculate the lowest fractional
population of foreground red clump stars that can provide
at least half of the observed optical depth for various limiting 
parameter cases (\eg how precisely do we need to exclude a foreground
population to exclude it as a major source of microlenses). 
For example, with $\Gamma = -4$, and the other
parameters
unchanged from those adopted by Bennett, the foreground population needs
only to be 0.9\% of the red clump stars. Therefore, even if 87\% of
the VRC stars are due to another phenomenon (\eg stellar evolution),
there are still sufficient foreground stars to account for half of the
observed optical depth. 
For $\Gamma = -2$, $m_1 = 0.1$, and $m_{to} = 1$, only 45\% of the
VRC stars need be foreground stars for $\tau_{fg} = 0.5 \tau_{\mu}$.
It will be exceedingly difficult to exclude 
such a small fraction of foreground stars without direct distance
measurements to the lenses. The more extreme values of $\Gamma$ are 
almost certainly inappropriate over a large mass range (for example, they
would predict very large $M/L$), but the appropriate value is not
well determined.

\subsubsection{Gould 1998}

Gould examined whether ZL's implied $M/L$'s, both for the LMC disk
and the foreground population are reasonable and whether there
is any existing evidence for tidal streamers/tails superposed on 
the LMC. The only existing photometric survey that is relatively
deep over a sufficiently 
large area is the photographic survey 
of de Vaucouleurs (1957). Based on
the surface brightnesses measured in that survey and the surface
mass density necessary to reproduce the observed lensing optical depth, 
Gould concluded that the implied $M/L$ of 12 was unacceptably high (if
one was to account for the lensing by appealing to a stellar population that
resembles known stellar populations). Furthermore, he concluded that
de Vaucouleurs's maps shows no evidence of a component (with 
sufficient surface mass density) extending beyond
the LMC and concluded that either (1) there is no such component, (2)
it is concentrated on the LMC and has a surface brightness profile
similar to the LMCs, or (3) it is uniform over the entire area of
the survey and so, much more extended than the LMC. 

In addition to the standard difficulties in low surface brightness
measurements with photographic material, the uneven foreground
emission is, according to de Vaucouleurs (1957), ``particularly
serious''. Because of ``galactic structures ... and the general
luminosity gradient toward lower latitudes'', de Vaucouleurs was
forced to apply a set of corrections that were constructed to
remove irregularities ``as well as possible by means of field
corrections'' and  also to apply ``considerable smoothing ... to
reduce spurious details.'' Because
of such difficulties, de Vaucouleurs's concluded that his procedure
``effectively excludes the very weak outer extensions''.
Given this irreversible treatment of the
data, it is difficult to ascertain the precision of the outer
surface brightness contours and the
reliability with which tidal features may be identified
(de Vaucouleurs used those data to
determine
a total magnitude  and a surface brightness profile for the LMC, so
his analysis did not require precise outer isophotes). Indeed, 
current studies are beginning to find signatures of possible stellar
tidal structures near the LMC (Majewski \etal 1998; Geisler \etal
1998). 

Gould compared the $M/L$ derived using
the mass density from ZL and the luminosity from de Vaucouleurs's
$V$-band measurement, converted to $R$ for an assumed $V-R = 0.5$,
to the $I-$band $M/L$ given by Binney and Tremaine (1987) for the
local Galactic disk (1.8 in solar units). We can redo that calculation
using our photometry to avoid the issues related 
to the photographic data and the assumed color conversions. Our
calculation has the added advantage that it provides the luminosity
at the position in the LMC being discussed.
We merge our data with
the deep HST luminosity function (available for $V$ and $I$ from 
Holtzman's data) and measure the $V$-band luminosity of the ZL
region to be $5.1\times10^7 L_\odot$ (for a distance modulus of 
18.5 and $M_{V,\odot} = 4.8$). This luminosity is a slight underestimate
because the HST data are incomplete at faint magnitudes ($M_V >
7$), but those stars contribute little to the total luminosity.
The corresponding average surface brightness of the ZL
region is 22.9 $L_\odot$pc$^{-2}$. 
The average
reddening, $E(B-V)$, measured for O and B stars in this area of the
LMC is 0.2 mag (Harris, Zaritsky, \& Thompson 1997) and
includes Galactic reddening. For a standard interstellar extinction
law (Schild 1977),
which is valid at optical wavelengths for dust in the LMC (Fitzpatrick 1985),
$A_V = 3.2E(B-V)$, 
so $E(B-V) = 0.20$ mag corresponds to $A_V = $ 0.64 mag.
After making this correction, the extinction-free V-band surface luminosity
is $41 L_\odot$pc$^{-2}$. This value is a slight overestimate if
the VRC stars are indeed in the foreground (because their dereddened luminosity
was included in the measurement), but the
foreground population is $<$ 10\% of the LMC population so the effect
is small. The ZL mass estimate, which corresponds to
103 $M_\odot$ pc$^{-2}$, implies $(M/L)_V = 2.5$.

We can also do the same calculation for
the $I-$band, which is less affected by dust.
The $I$-band luminosity of the ZL region (for stars with $I \le 19.0$)
is $3.2\times 10^7 L_\odot$. Again using Holtzman's data to correct
for missing fainter stars
and an $I$-band extinction correction of 0.38 mag,
the surface brightness is 32 $L_\odot$pc$^{-2}$ and 
$(M/L)_I = 3.2$. Finally, we note that the ZL region is not 
as luminous as other regions (\eg the left side of Figure 1) and
so the global $M/L$ for the LMC may be lower.

We compare these values of $M/L$ to measurements of $M/L$ for other galaxies.
The calculated $(M/L)_V$ is not beyond the range observed for 
spirals (cf. Kent 1986; although his values are for the $r$-band and some
slight color correction is necessary). 
The $I$-band values are easier to judge because many measurements
exist for either the local
disk ($(M/L)_I = 1.8$ as adopted by Gould from Binney \& Tremaine)
or for large samples of galaxies. Vogt (1995) presents 
$(M/L)_I$ for a wide range of spiral galaxies. Although Hubble types
Sd and Sm
are not well-represented in her sample,
we adopt the values for Sc types as appropriate (this is possible because
no strong type or luminosity dependence is found). The median
$(M/L)_I$ (converted to $H_0 = 75$ km sec$^{-1}$ Mpc$^{-1}$) is
2.5 and the upper quartile begins at $(M/L)_I = 3.1$. The median
value for Sb galaxies (appropriate for our galaxy) is 2.0, in
excellent agreement with the value of 1.8 from Binney \& Tremaine.
In comparison to these results, $(M/L)_I$ derived for the LMC
is slightly larger than expected, 
but not beyond the range of values observed
for other galaxies (it lies just inside the upper 25\%).
Given the reasonable agreement in the inferred LMC $M/L$ ratios
for both $V$ and $I$ and determinations
for other galaxies, and the 
uncertainties in the multiple steps
required for this comparison, we find no reason to conclude that ZL's
inferred LMC mass, and by association that of the foreground
component, can only be explained with a dark matter dominated (\eg M/L $>$
12)
stellar population. Again, we stress that ``factor
of two'' effects are ubiquitous and can work in either direction (for
example, ZL adopted a disk inclination of 33$^\circ$ for the LMC
but published values range from 22$^\circ$ (Kim \etal 1998) 
to $48^\circ$ (Bothun \& Thompson 1988) which result in velocity
inclination corrections that vary by a factor of two). We conclude
that there are large uncertainties in the derived masses, but 
that masses as large as implied by ZL's analysis are not
sufficiently extreme to be confidently excluded.

Finally, Gould noted that ZL exclude the contribution of the dark
matter in the LMC to the rotation curve 
and thereby inflate the stellar $M/L$ (and so overestimate the
lensing contribution of a standard stellar population). Kim \etal (1998)
recently published and analyzed a high resolution H I map of the
LMC. They present mass models and conclude that no
dark matter halo is required to fit the H I data. To fit their H I observations 
{\it and} the carbon star data at larger radii (Kunkel \etal 1997), they do
use a model that includes a dark matter halo. From their
Figure 6, we infer that the halo's contribution to the mass
inside 3 kpc is $<$ 20\%. Therefore, although while Gould's 
assertion is strictly true, the discrepancy is apparently minor given
the ``factor-of-two'' uncertainties present elsewhere.
For their global model,
Kim \etal calculate that the stellar $(M/L)_R$ is 1.8
(where the luminosity is taken from de Vaucouleurs (1957)), which is less
than a factor of two discrepant with either the $V$ or $I$-band
$M/L$'s derived above.
Without a dark matter halo (or with a smaller contribution from
the dark matter halo within 4 kpc), the stellar $M/L$ would
increase by up to $\sim 20\%$ and be in even better agreement with our
estimates of $M/L$.

\subsubsection{Johnston 1998}

ZL suggested that one possible source of foreground stars was 
the suggested
tidal streamer (Zhao 1998a) that arose as a result of the LMC-Milky Way
interaction in a previous perigalacticon passage.
Johnston examined this suggestion in greater detail
and concluded that either such a feature would lead
to unacceptably high faint star counts across a large section
of the sky or that the feature would
disperse quickly ($\sim 10^8$ years).
Her study does not exclude other possible sources of foreground
stars such as tidal material 
from an SMC-LMC interaction or from a denser than
expected LMC halo. 
The SMC-LMC interaction that is thought to have occurred within
the last few$\times 10^8$ years (Gardiner \& Noguchi 1996)
may be an attractive candidate for the source
of tidal material along the line of sight. The environment
around the LMC-SMC is sufficiently complicated (\cf Putnam \etal 1998)
that the presence of stars outside the body of the LMC with a total
mass of a few percent of the
LMC should not be surprising.

\subsubsection{Beaulieu and Sackett 1998}

The study by Beaulieu and Sackett
provides the strongest challenge yet to the interpretation
of the VRC as a foreground population by suggesting that the young
($< 1$ Gyr) LMC stars evolve into a superluminous red clump phase that
populates the VRC region of the CMD. This argument is bolstered by
theoretical isochrones for such stars, the clear presence of young
main sequence stars in the LMC, the identification of 
such superluminous red clump stars in the Galactic HIPPARCOS data,
and the identification of such stars in the Fornax and Sextans A
dwarf galaxies (Stetson 1997; Dohm-Palmer \etal 1997). 
Quantitatively, they support their
suggestion by presenting models for the star formation history of the
LMC that produce the observed number of stars in the VRC region of 
the CMD. However, recall from \S3.1.3 that to the limit of our current
understanding of the IMF, it is possible that only a minor fraction ($<50$\%)
of the VRC stars need be foreground stars to account for a large
fraction of $\tau_\mu$. Therefore, a precise estimate of the fraction
of VRC stars that are young red clump stars is necessary.

If the star formation history adopted by Beaulieu and Sackett 
is correct, then their models
predict that any foreground population contributes at most only
slightly ($< 25$\%) to the VRC. However, once again the models depend on 
assumption regarding complex issues such as the star formation history
and IMF. We examine the dependence of those models on some of 
the selected parameters. 

We adopt the Bertelli \etal (1994) isochrones, a standard (Salpeter) IMF 
slope, and add observational uncertainties to generate CMDs
corresponding to a variety of star formation histories. From these
data we have recreated the histogram of differential magnitude 
relative to the red clump centroid for the narrow color cut around
the RC that was
presented by ZL. We have selected three star formation histories for
comparison:
(1) a constant star formation rate (SFR) over the most recent 1 Gyr,
and nothing prior, (2) a
constant SFR over the most recent 1.3 Gyr and nothing prior, and (3)
the star formation history presented by Geha \etal (1997)
based on analysis of deep HST CMDs. 

There are two representative 
tests of the models. First, the ratio of the number of 
main sequence stars with 
main sequence lifetimes $\sim 0.8$ Gyr to the number of 
VRC stars constrains whether the model is able
to generate VRC stars provided that there has been 
a recent episode of star formation. This 
ratio was used by Beaulieu and Sackett to demonstrate the feasibility
of their hypothesis. Second, the ratio of VRC to RC stars tests
whether the entire star formation history is consistent with
the observed number of VRC stars relative to RC stars. 
If the first test is not
satisfied, then one can conclude that this phase of stellar
evolution (young, massive RC stars) 
cannot populate the VRC in sufficient numbers, regardless of the star
formation history.
If the first test is
satisfied,
but the second is not, then one concludes that this
phase of stellar
evolution can populate the VRC in sufficient numbers, but that
the recent star formation rate 
was insufficient to account for the observed
number of VRC stars. 
Beaulieu and Sackett demonstrated that the first test is satisfied
(their models were able to reproduce between 75\% and 100\%
of the ratio of main sequence
stars to VRC stars).

We present the histograms from vertical (constant color)
cuts through the simulated 
CMDs at the position of the red clump in Figure 6.
Our first conclusion is that stellar evolution is able, in principle,
to populate the VRC. Our second
conclusion is that a model with a constant star formation over the last
1.3 Gyr (but no previous star formation) is at least one 
model that
reproduces the observed ratio of VRC to RC stars. This conclusion
confirms
the result of Beaulieu and Sackett that a model
with a constant SFR over the last $\sim$ Gyr is able to generate
the appropriate number of VRC stars. Our third conclusion is that
a model with an LMC star formation history independently derived from
HST CMDs will significantly {\it underproduce} the
number of VRC stars
(defined to have $-1.4 < \Delta m_V < -0.7$)
relative to the RC stars (defined to have
$-0.325 < \Delta m_V < 0.325$) within this color cut (for the Geha \etal 
star formation history, the calculated ratio is only 26\% of the
observed ratio).
We conclude that given the uncertainties in
the star formation history, stellar models
(\eg overshooting which affects the lifetime),
and binarity (which we did not explore, but which may provide
additional free parameters),
we are unable to unequivocally determine
whether stellar evolution does or does not populate the VRC region
of CMDs in sufficient numbers, but that a model with a ``realistic''
star formation history defined independently of the VRC
falls short by about a factor of two. 
Our understanding of the star formation history
of the LMC is inadequate to be able to reach definitive conclusions 
about the relative fraction of young red clump stars that
constitute the VRC with better than a 50\% uncertainty. We will
attempt to constrain this hypothesis with our new data in \S 4.

\subsection{Summary of Previous Studies}

We close this section by stressing that while certain ``reasonable''
assumptions can lead one to conclude that foreground stars are
at most a small fraction of the VRC stars or that even if most
of the VRC stars are foreground RC stars
this population will not contribute significantly to
$\tau_{\mu}$, these conclusions depend sensitively on adopted
assumptions. {\it A foreground population
that is responsible for a significant fraction of $\tau_{\mu}$ is
consistent with the available data.} If it is so difficult to exclude
a relatively well-defined, spatially extended stellar population, it will be
nearly impossible to exclude more subtle populations such as a slightly
thicker LMC (Graff \etal 1998) or a population less than 10 kpc behind
the LMC (Zhao 1998b) without having direct distance measurements
to the lenses. 

\section{New Constraints on the Nature of the VRC}

We set out to further constrain the nature of the VRC stars in two 
ways: (1) by obtaining stellar photometry over a larger area of the LMC
to examine the distribution of VRC stars relative to other LMC stellar
populations, and (2) by measuring VRC and RC radial velocities to place
kinematic constraints on the population. The data acquisition 
was discussed in \S2, we now discuss the results from
each of those datasets below.

\subsection{Spatial Distribution}

The spatial distribution of the VRC stars can provide a fundamental
clue to the nature of the VRC stars. If the VRC stars are seen 
exclusively in sites of recent ($\ltsim 1$ Gyr) star formation, then
the Beaulieu and Sackett model would be
confirmed. Alternatively,
if VRC stars are found in regions where there are no comparably aged stars, 
then VRC stars are not exclusively young red clump stars.
Therefore, a simple test is to compare the spatial distribution
of VRC stars and main sequence stars that are of approximately the same
mass as the VRC progenitors.

The VRC stars here are defined in color as in ZL and in magnitude
as by Beaulieu \& Sackett for more direct comparison (the same color
criteria as Beaulieu \& Sackett applied
could not be used because we do not have $R$-band data). 
We define the VRC stars to have $2.85 < C < 3.57$ and $17.8 \le V \le 18.5$.
We define the main sequence (MS) stars to which we compare 
to have $-0.2 < B-V < 0.3$ and $19.15 < V < 19.5$ (the same magnitude
range as chosen by Beaulieu \& Sackett). The different color criteria
between our definitions and those of Beaulieu \& Sackett have a 
minimal effect. For example, their preferred model results in a
ratio of VRC stars to MS stars of 0.23 for their selection
parameters and our comparable model
(constant star formation for the last 1.3 Gyr and no star formation
prior to that) results in a 
ratio of 0.20 for our selection parameters.
We bin our data into $6^\prime \times 6^\prime$ squares 
(comparable to the area of
a single image in the Beaulieu and Sackett study). 

In Figure 7 we compare the stellar density distributions of VRC and 
MS stars. For the assumption adopted by Beaulieu \& Sackett of 
constant star formation rate over the brief time separating the
formation of the VRC progenitor stars and the current MS stars, 
the ratio of VRC to MS stars should be a constant across the area. 
From the Figure
it is evident that there is some correlation between the two (\eg
the enhancement of VRC stars in the lower left and upper right of the 
area), but the VRC distribution appears more uniform than 
the MS distribution.
For a quantitative comparison we divide the VRC image by the
MS image and examine this ratio on a pixel-by-pixel basis. The
distribution of those ratios as a function of MS stellar density is 
plotted in Figure 8. 

The systematic variation of the mean ratio with MS star density indicates that
the number of VRC stars is not directly proportional to the number of MS
stars. Variations of this ratio could be due to a variation of 
the star formation rate as a function of location over the
time in which the current 
VRC stars were formed and the current MS stars were formed, to
the greater diffusion of the slightly older VRC stars from their star
formation sites, or to the presence of 
an additional component in the VRC that is uncorrelated with the MS
stars. Nevertheless,
variations seen in Figures 7 and 8 demonstrate (1) that 
the measured ratio in any single $6^\prime
\times 6^\prime$ box cannot be directly compared to the ratio
calculated from simulated CMDs and (2) that at least some VRC stars
are associated with recent star formation regions.

Regardless of the correct explanation for the 
variation in VRC/MS, this ratio
must approach that estimated from the synthetic CMDs
when measured over a large region of the LMC if the young RC
hypothesis
is correct.
The average of VRC/MS over our entire field
is 0.31, while the models (using either
the model with constant star formation over the last 1.3 Gyr or
the Geha \etal star formation history) predict 0.20 and 0.14,
respectively. The comparison to the ``realistic'' star formation
history model suggests that over the entire area $\sim$50\% of the VRC
stars are not related to young red clump stars and that the
behavior of VRC/MS is not entirely attributable to either of the
first two possibilities. 

The existence of a uniform VRC component
can be quantified using a simple model for
the ratio of VRC stars to MS stars. If the young red clump component
of the VRC stars is an unknown fraction, $f$, of the MS stars and
the remainder is in a uniform component of $A$ stars per bin, then
\begin{equation}
{VRC\over MS} = {A + f \cdot MS\over MS}.
\end{equation}
Fitting this equation to the data presented in Figure 8, we derive
$f = 0.088$ and $A$ = 7.25, with $\chi^2 = 1.2$. 
This result implies that 28\% of the VRC
stars are young red clump stars and that 
the remainder  is in a uniform component,
which {\it could} be a foreground population. 
Alternatively, one can take $f$ as given from stellar population
models. For example,
from the Geha \etal star formation history, $f= 0.14$. Fitting such
a model we derive $A$ = 4.88 with $\chi^2 = 2.9$. While the
latter is a poorer fit (which can be excluded with 90\% confidence),
it still predicts that only 45\% of the VRC stars are young red clump
stars. Models that predict that a larger fraction of VRC stars are
young red clump stars can be discriminated with higher confidence.

As discussed above, the possibility that VRC and MS stars have
different clustering scales can affect this analysis.
Variable clustering scale may be present if the young red clump stars have had
sufficient time to
diffuse farther away from sites of star formation
on average than have the MS stars. To check if this is a substantial
contributor to the previous result,
we redo the analysis 
using boxes that are $3^\prime \times 3 ^\prime$ and 
$12^\prime \times 12 ^\prime$. The 
results are shown in Figure 9 and are consistent with the previous
best fit relation. The best fit model to the data from the
$3^\prime \times 3 ^\prime$ boxes
suggests that 21\% of the VRC
stars are associated with the MS stars (instead of 28\% as derived
from the best fit to the data from the $6^\prime \times 6^\prime$ 
boxes). This difference is in the sense expected if the
VRC and MS stars are distributed similarly but the VRC stars
have slightly larger clustering scales. Nevertheless, this difference
does not affect the conclusion that a large fraction of the
VRC stars are unassociated with the MS stars.
The distribution of VRC/MS from boxes ranging
from 3$^\prime$ to 12$^\prime$ on a side 
give consistent results that
a minority fraction ($<$50\%) of the VRC stars are directly correlated with the
main sequence stars. Although the simplest interpretation of the 
spatial modeling is consistent with the results from the simulated
CMDs, other interpretations involving variable star formation or
more complex dynamical evolution are possible.

Why does our conclusion regarding the origin of the VRC differ from
that of Beaulieu \& Sackett? 
Over a wide range of MS star densities the
ratio of VRC/MS in 6$^\prime \times 6^\prime$ areas 
is $\sim 0.2$. It is therefore not unexpected that when
Beaulieu and Sackett analyzed four fields of this size, three of those
had ratios in that range.
On average (weighing their four fields equally), their measurement of
VRC/MS is 0.33, which is nearly identical to our ``global'' average
of 0.31. 
Therefore, the different conclusions are not due to differences in
the definition of the VRC or MS stars (or to gross differences
arising from reddening or completeness corrections), 
but rather in our ability 
to explore a wide range of MS star densities, average over many
fields to improve our statistics, and on our use of the Geha \etal
star formation history. Clearly, the range of possible star formation
histories is infinite and it is quite possible that there are
many plausible histories that produce a sufficient number of
young red clump stars, but the detection of dark matter should
not rest on such ill-constrained models.

\subsection{Radial Velocities}

The radial velocity distribution of VRC stars is shown
in Figure 10. The small number of stars at $\sim -275$ km sec$^{-1}$
are Galactic stars and they illustrate that even though
the location of the VRC in the CMD places it near the locus of 
Galactic contamination (disk stars), the level of contamination is 
minor. The bulk of the VRC stars
are centered at 0 km sec$^{-1}$ rather than $275$ km sec$^{-1}$,
which is an artifact of how the velocity template was generated. 
The velocities for the red clump stars are presented in 
Figure 11, and the similar centroid velocities for the
VRC and RC demonstrate that the VRC stars have an average velocity 
that is statistically indistinguishable from that 
of the LMC. This result confirms
the result of Ibata, Lewis, \& Beaulieu (1998) based on 16 VRC stars.

As Ibata \etal argue, the agreement between the VRC and LMC
velocities argues for a close relationship between the VRC and LMC
stars and almost certainly excludes the possibility that the VRC is
due to a foreground population unrelated to the LMC (such as an
intervening galaxy or tidal streamer). Therefore, any 
interpretation of the the VRC as foreground material,
is now constrained to place
these stars in close association with the LMC (they must be
either gravitationally bound or recently tidally removed).

We proceed to measure the velocity dispersion of the VRC stars
in two ways. The first approach we use is 
an iterative maximum likelihood fitting
algorithm that fits to the unbinned distribution of data. 
We have eliminated stars with $|\Delta v| > 100$ km sec$^{-1}$
and with $R < 6$. These cuts eliminate galactic contamination and
unreliable velocity measurements. The results from the remaining 190
VRC stars are $\bar v = 1.2 \pm 3.2$ km sec$^{-1}$ and $\sigma = 18.4 \pm
2.8$ km sec$^{-1}$ (95\% confidence interval used throughout for
uncertainties in mean velocities and velocity dispersions). 
The second approach we use fits a Gaussian to
the binned data.
By convolving the underlying distribution, characterized
by a mean velocity, $\bar v$, and dispersion, $\sigma$, with the
distribution of observational errors, we generate the expected
distribution of radial velocities. With a standard definition of $\chi^2$ 
(the sum of the differences squared 
between the histogram and the convolved expectation divided by the
uncertainties from Poisson statistics squared),
we associate a $\chi^2$ with each model and determine
the best fitting model and allowed parameter ranges.
For the same sample of stars as used for the unbinned parameter
fitting, and a bin size of 10 km sec$^{-1}$ bins
we find $\bar v = 3.2_{-4.4}^{+1.8}$ km sec$^{-1}$ and
$\sigma = 17.5_{-3.7}^{+3.8}$ km sec$^{-1}$ ($\chi_2 = 1.1$). The two approaches provide a consistency check on the
algorithms used to determine the underlying dispersion given the
substantial observational uncertainties. However, we adopt the
results from the first method because the results do not depend on 
an adopted bin size.
As is evident from Figure 10, the Gaussian model provides an acceptable
fit. Finally, we attempt to correct for the fact that the two fields,
even though they are close to each other, could have a different mean
velocity due to the projection of the LMC's rotation curve on the
plane
of the sky. Rather than rely on a model, we make the most favorable
assumption for lowering the velocity dispersion, which is to set the
mean velocities of stars in the two VRC fields equal. The resulting velocity
distribution has $\sigma = 18.0 \pm 2.8$ km sec$^{-1}$.
We conclude that the measured VRC
line-of-sight velocity dispersion is $>$ 15 km sec$^{-1}$ with 
95\% confidence. 

For the red clump stars (Figure 11), we have larger observational uncertainties
and fewer data (75 stars), so the kinematic constraints are less precise. 
The unbinned fitting results in $\bar v = -4.6 \pm 9.0$ km sec$^{-1}$
and $\sigma = 32.2 \pm 7.6$ km sec$^{-1}$.
The mean velocity of the fitted Gaussian to the
binned data is $-6_{-12}^{+21}$ km sec$^{-1}$ and
the dispersion is 32$^{+19}_{-16}$ km sec$^{-1}$. 
The mean velocity calculated directly from the data
(no assumed model) is $-4.6\pm9.0$ km
sec$^{-1}$ (1$\sigma$), which is in exact agreement with the parameters
from the unbinned fitting algorithm. 

We use the velocity dispersion vs. age relation 
for stellar populations given by Villumsen (1985),
\begin{equation}
\sigma_z = \sigma_{z,0}(1 + t/\tau)^{0.31},
\end{equation}
where $\sigma_{z,0}$ is the velocity dispersion of the cold gas,
$t$ is the age of the population, and $\tau$ is a ``diffusion''
timescale to estimate a mean ``kinematic'' age for the VRC stars. 
This relation was derived for Galactic stars. While we
assume that the general physical behavior is the same, we 
normalize the relation for LMC populations.
Adopting an initially vertical velocity
dispersion of 5.4 km sec$^{-1}$ (from the H I gas) and 
19.1 km sec$^{-1}$ for a population of planetary
nebulae of mean age 3.5 Gyr (Meatheringham \etal 1988), we derive $\tau
= 6.0 \times 10^7$ yrs, in close agreement with the value derived
by Wielen (1977) of $5\times 10^7$ yrs for a growth
rate of $(1+t/\tau)^{1/2}$ (for Galactic stars). 
Using the Villumsen age-velocity
dispersion, we calculate that the mean age of the VRC population
is $3.1 \times 10^9$ years and that a 95\% confidence lower limit on
the age is
$1.8 \times 10^9$ years (this estimate is only valid if
the VRC stars are a single, disk population). Similar value are obtained
using the original Spitzer and Schwarschild (1951,1953) formula
($\sigma \propto (1+t/\tau)^{1/3}$) or Wielen's parameterization.

Does the age-velocity relationship provide reasonable and reliable
estimates of the velocity dispersion for young stellar populations? 
To test whether young stellar populations distributed through the
LMC do have a low line-of-sight ($<$ 15 km sec$^{-1}$) velocity dispersion
we reanalyzed the velocity data obtained by Feast, Thackeray, \&
Wesselink (1960,1961) on supergiants in the LMC.
We adopt the same sample of stars that they use in their dynamical
analysis. Because these stars, which are
typically O through A supergiants, are 
distributed over the entire LMC, 
the velocities have to be corrected for the rotation of the LMC. 
We take the simplest model (a flat disk with a constant rotation curve at all
radii) to avoid using additional parameters to
deflate the velocity dispersion, and use the mean inclination
and semimajor position angle found by 
Kim \etal (1998) from the study of the H I kinematics ($i = 33^\circ$,
a semimajor position angle of $-10^\circ$).
We fit the model to derive the
rotation velocity,
42 km sec$^{-1}$, and the systematic velocity, 271 km sec$^{-1}$).
We then subtract this model from the data and find the velocity
dispersion to be $19.6 \pm 4.4$ km sec$^{-1}$. However, three of the
most deviant points from the mean rotation curve are at radii $< 1$ kpc
for which the rotation curve is steeply rising (Kim \etal) and where
the influence of the bar is probably not negligible. Three other
stars are also on the extreme tails of the distribution and appear
to artificially inflate the dispersion. The observed
distribution 
does not appear to be Gaussian and a model distribution with the
calculated dispersion
overpredicts the number of stars at intermediate velocity
differences and underpredicts the number at small velocity
differences (see Figure 12). Eliminating the three
stars near the center of the LMC and the three most extreme outliers 
leaves us with a distribution that is well fit by 
a Gaussian of velocity dispersion $10.0 \pm 3.8$ km sec$^{-1}$ (note
that the VRC velocity distribution shows no sign of extended
non-Gaussian wings and so its dispersion does not appear
to be inflated by outliers). 
Removing only the two stars at $\Delta v = 50$ km sec$^{-1}$
results in $\sigma = 12.5 \pm 3.6$ km sec$^{-1}$, which is still inconsistent
with the measured VRC dispersion at the greater than 95\% confidence
level, but roughly the value expected for a 0.8 Gyr old population. 
Finally, we stress that the rotation model is highly simplified
and that deviations, such as a systematic change in the line of nodes,
warping, and a non-flat rotation curve 
have been observed (Kim \etal) and will cause us to
overestimate the dispersion of these stars.
We conclude that there are stellar populations that are
sufficiently young and dynamically cold 
($\sigma \sim 10$ km sec$^{-1}$) and
that such small velocity dispersions
can be measured using velocity measurements that have
uncertainties that are $\sim 10$ km sec$^{-1}$.

We can further attempt to constrain the VRC population by modeling
the velocity distribution as a sum of two Gaussians, one with the 
velocity dispersion expected for a population of age $\sim 0.8$ Gyr and
the other undetermined. By fitting the range of relative
amplitudes, we can place a limit on the fraction of the VRC
population
that could be in a cold, young disk component. We fix the
velocity dispersion of the young component at 12.3 km sec$^{-1}$
(derived from the Villumsen dispersion-age relation), and allow 
the means of the two Gaussian to vary. We find that models in
which the young disk component contributes more than 77\% of all
the VRC stars can be excluded at $>$95\% confidence (all models
where the young population contributes $< 77\%$ are allowed, and
for models with a high fraction of young stars 
the velocity dispersion of the undetermined component is 
$\sim 35 \pm 10$ km sec$^{-1}$). 
While these results do not provide as strong a constraint as that
provided by the 
spatial distribution, it is entirely consistent with the
implications from the stellar population models and the spatial
distribution that the young red clump stars contribute $\ltsim$ 40\% 
of the VRC stars.

The VRC stars have a lower velocity dispersion than the RC stars, and
certainly lower than one would expect for a halo population of the 
LMC. Therefore, at first glance 
it appears that the VRC stars could not be a constituent
of a dynamically hot component. However, we caution that because of
the magnitude selection of VRC stars (\ie that they be 
brighter than the RC stars) we would have selected only those stars on the
near side of the LMC ``halo''.
The measured velocities may not be a fair
sampling of the velocity distribution of these stars and could produce
a significant underestimate of the true velocity dispersion,
if indeed they are distributed throughout the LMC halo.

There are at least two caveats with regard to
this velocity analysis. First,
velocity uncertainties are notoriously hard to estimate
precisely. Because the measured velocity dispersion is not grossly
larger than the velocity uncertainties, the resulting dispersion
is {\it highly} sensitive to the adopted uncertainties. It is possible that
the velocity dispersion is inflated by observational uncertainties.
An increase in the velocity uncertainties by 50\% lowers the best fit
velocity dispersion to 12.8 km sec$^{-1}$ and the 95\%
confidence limit on the velocity dispersion to 9.2 km sec$^{-1}$ (at
which point a young population could not be ruled out).
Although we have no evidence that indicates that the velocity 
uncertainties are severely underestimated and our
dispersion measurement is consistent with that of Ibata, Lewis,
\& Beaulieu (1998), the possibility exists.
A second caveat is that giant branch stars may contaminate the VRC region. 
In particular, some contamination may be due to stars on evolutionary
blueward loops (``blue noses'') near the AGB bump (Gallart 1998).
Such stars are older than 1 Gyr and would have the necessary kinematics and
uniform spatial distribution that are observed. 
Therefore, advocates of halo MACHOs can
assert that the VRC consists of young red clump stars plus an older
contaminating component, such as blue nose stars. We conclude that
the composition of the VRC is mixed, but that a significant intervening
population has not yet been ruled out.

\subsection{Microlensing Revisited}

In this discussion, we have suggested that some 
adjustments to the parameters
adopted
by ZL in their calculation of the optical depth might be warranted. 
For example,
as many as half of the VRC stars may be young red clump
stars rather than foreground stars. How does this affect the possible
microlensing optical depth from such a population? 

A straightforward
calculation, along the lines of Gould (1998) goes as follows.
In the ZL area there are 2471 VRC stars / sq. degree, the ratio
of VRC to RC stars is 0.08 (as defined in \S3.1.3), 
the ZL area is 2.93 sq. degrees, and the total $I$-band luminosity is
$7.1 \times 10^7 L_\odot$ (corrected for extinction and stars fainter
than
the magnitude limit of the
MCPS, \cf \S3.1.4). We presume that half of the VRC 
stars are foreground stars. Assuming that the foreground stellar population
is identical to the LMC stellar population, the luminosity of the foreground
population is $2.8\times 10^{6} L_\odot$. For a typical $(M/L)_I$
for Sc galaxies (2.5; Vogt 1995), the total mass in this component
over the ZL area of the LMC
is then $7.1 \times 10^6 M_\odot$. The surface mass density is then
$2.4 \times 10^6 M_\odot$ per sq. degree, or 5.0 $M_\odot$ pc$^{-2}$
if the foreground stars are at 40 kpc. Using the same formula
as used by Gould (rewritten for $\tau_{fg}$), 
\begin{equation}
\tau_{fg} = {2.9\times10^{-7}\over 47} ({D \over 10 {\rm kpc}})
\end{equation}
where $D \equiv d_{ol}d_{ls}/d_{os}$ and $d_{ol}$, $d_{ls},$ and
$d_{os}$ are the
distances between the observer, the lensing structure, and the sources
in the LMC, we calculate that $\tau_{fg} = 2.5\times10^{-8}$  
($\sim$ 9\% of $\tau_{\mu}$). This result agrees with 
Bennett's (1998) and  Gould's (1998) conclusions 
that even if there is a foreground
population it would not contribute significantly to $\tau_\mu$.

However, to demonstrate the effect of various uncertainties, we redo the
calculation with slightly different (but plausible) parameters. First,
we calculate the VRC/RC ratio using our entire dataset and obtain
that the ratio is 0.096. 
Instead of adopting that 50\% of the VRC is foreground, we will
use the result from the best fit model described in \S4.1 that
72\% of the VRC is in a uniform component, which we will attribute to a 
foreground population. We will also use $(M/L)_I = 3.2$, which 
is consistent
with what ZL derived and within the observed range of 
$(M/L)_I$'s for Sc galaxies (Vogt 1995), and a distance to the lenses
of 35 kpc. For these parameters, we 
derive that $\tau_{fg} = 9.2\times 10^{-8}$, which is 32\% of 
$\tau_{\mu}$. If $\tau_{\mu}$ is revised downward to $\sim 2 \times
10^{-7}$ (Bennett 1998), then this population can account for about
50\% of the observed microlensing. 

Finally, we stress that the foreground population does not need to produce
$>$50\% of $\tau_\mu$ to be important. A foreground
tidal or halo stellar population
should have a counterpart behind the LMC. The lensing of that
population by LMC stars will result in a comparable
contribution to
$\tau_\mu$ (Zhao 1998b). In addition, if the Galactic disk contributes
an optical depth as large as 
$1 \times 10^{-7}$ (Evans \etal 1998), then the VRC
foreground
population only needs to contribute 0.25$\tau_\mu$ for these three populations
to account for $\tau_\mu$. 
Finally, the ``standard'' intervening stellar components (as estimated by
Alcock \etal 1997a) have $\tau = 0.5\times 10^{-7}$, with significant 
uncertainties (\cf Aubourg \etal 1999) even prior to appealing to unknown tidal tails or tilted
Galactic disks. 
Therefore, arguments against 
the importance of intervening populations on the interpretation of
microlensing events should not consider a single population
in isolation, but rather the sum of all plausible populations.

\section{Summary}

The microlensing surveys have convincingly demonstrated that 
microlensing can be detected. When the number of lensing events
observed along the line of sight to the Large Magellanic Cloud (LMC)
is compared to that expected in
the standard model of the structure of the LMC and the Galaxy 
(Alcock \etal 1997a) there is an excess of lensing events. 
Several authors, including
some of the authors of this paper, have suggested previously 
that the ``excess'' lenses may not be halo MACHOs but rather 
normal stars along the line of sight. In particular, Zaritsky \& Lin
argued that the color-magnitude diagram of the LMC has a vertical 
extension of the red clump (the VRC) that is consistent with the presence
of a foreground population that lies within 15 kpc of the LMC. 
Various studies have attempted to refute almost
every aspect of that claim. 

We adopt a two step approach in revisiting this issue. 
First, we examine each of 
the rebuttal studies
to determine whether the arguments presented can
convincingly discriminate
against the possible foreground population suggested by ZL. 
Second, we obtain new data to test the
most damaging of the rebuttal studies (Beaulieu \& Sackett 1998), which
suggested that the VRC stars are all young, massive red clump stars.

From our examination of the published studies,  
we find that to within the observational and model 
uncertainties one cannot convincingly refute (or prove) the existence of
a foreground population that exists within 0-15 kpc of the
LMC along the line of sight and that provides a significant
($\sim 50$\%) fraction of the total microlensing optical depth.
Therefore, we conclude that systematic uncertainties dominate
the current efforts to measure the total mass in halo MACHOs. Additional
lensing events, if they do not (1) dramatically alter the measured
microlensing optical depth, (2) dramatically alter the spatial 
distribution of lenses across the LMC, or (3) provide a large
number of anomalous lensing events for which transverse velocities can
be calculated (Alfonso \etal 1998; Alcock \etal 1998, Rhie \etal
1999), will not allow us to 
reach a significantly
different conclusion about the nature of the lenses. 
However, even if an intervening stellar population is responsible for
a significant fraction of the microlensing events, some events
may be due to halo MACHOS and these lenses could still be
a fundamental component of the Galaxy. Direct distance estimates
for lenses
will eventually avoid the complicated issues described here 
in understanding the line of sight stellar distribution.

From our new data and analysis, we find three reasons to question
the
suggestion that was quantitatively pursued by Beaulieu
\& Sackett (1998), that the VRC stars are exclusively young, massive red clump
stars.
First, synthetic color-magnitude diagrams created using a star
formation history derived independently 
from deep HST data (Geha \etal 1998) 
suggest that $<$ 50\% of the vertical red clump (VRC) stars are young,
massive red clump stars. Second, a detailed comparison of
the VRC and young main sequence stars,
suggests that $\ltsim 40$\% of the VRC stars correlate spatially
with the MS stars.
The remainder are
in a more uniform distribution, and so are unlikely to be young red clump
stars. 
Lastly, our measurement of the velocity dispersion of 190 VRC stars
($18.4\pm 2.8$ km sec$^{-1}$ (95\% confidence interval) is
inconsistent with the expectation for a young $(\ltsim 0.8$ Gyr) 
disk population (but consistent with that of the red giants). 
We confirm the measurement 
by Ibata \etal and agree with
their conclusion, that even if the VRC stars are foreground
stars they are unlikely to be from a stellar population unassociated
with the LMC. 
We conclude that various arguments suggest that
$\sim 40$\% of the VRC stars are young,
massive red clump stars. The remaining stars may be attributable 
to another stellar
evolutionary phase or to an intervening population associated
with the LMC (possibly an extended halo, stars that were recently tidally
removed, or possibly a much thicker, more massive disk than previously expected
(Aubourg \etal 1999)).

We apply a straightforward calculation for the microlensing optical
depth that avoids some of the subtle issues of star formation
histories
and IMFs by relying only an adopted $M/L$ for the stellar populations,
which we determine empirically from the studies of other galaxies.
We find that for one set of parameters the optical depth from the
possible foreground population (consisting of $\sim$  50\% of the
VRC stars) is $<$ 10\% of the optical depth measured by Alcock \etal
(1997). For somewhat different parameters, we find that the
microlensing
optical depth from the allowed foreground population is $\sim$ 30\%
of the Alcock \etal optical depth. 
More extreme, but not yet excluded, parameter
choices can lead to percentages $>50$\%. In combination with
known stellar components and other possible sources of contamination
(\eg a background population, a slightly thicker LMC, and more Galactic disk
stars), this population can produce $\tau_\mu$. Whether it does, is 
an open question.

We have, given our stated motivation of strongly questioning whether
the data yet demand the need for halo MACHOS, focused on attempting
to remove published objections to the intervening stellar 
population model. Some intervening populations, such as
an intervening stellar stream or dwarf galaxy have by now
been excluded (cf. Alcock \etal 1997b and Ibata \etal 1998). 
We have argued here that a population associated with the 
LMC (either within the LMC halo or very recently tidally lost)
has not yet been excluded.
In the interest of some balance, we restate the
basic arguments against interpreting the VRC as an intervening population.
First, stellar evolution models and plausible star formation
histories are able to account for the number of VRC stars. 
Second, the measured kinematics of VRC stars do not show any grossly
anomalous behavior (a difference in mean velocity with the LMC
or non-Gaussian behavior) and therefore do not argue strongly
against the VRC stars being in the LMC disk. Lastly, 
even if one accepts the
interpretation of VRC stars as foreground ({\it fg}), 
plausible models for the conversion
of the number of VRC stars into $\tau_{fg}$ result in $\tau_{fg}
\ll \tau_{\mu}$. However, our goal was to demonstrate that as
reasonable (and perhaps correct) as these arguments may be, 
numerous large uncertainties remain. Because of the
importance of the nature of dark matter to a wide range
of astrophysics, significant uncertainties must be removed before
the case for its identification becomes compelling.

In conclusion, despite various investigations into the nature of
the VRC population (including this one), we are only able to reiterate
the closing sentence of ZL's abstract --- ``We conclude that the
standard assumption of a smoothly distributed
halo population out to the LMC cannot be substantiated without
at least a detailed understanding of several of the following:
red clump stellar evolution, binary fractions, binary mass ratios,
the spatial correlation of stars within the LMC, possible variations
in the stellar populations of satellite galaxies, and differential
reddening --- all of these are highly complex.'' To calculate the
microlensing optical depth from such a population one must
add to that 
list of problem issues the initial mass function and star formation history. 
Given these uncertainties, we find that the hypothesis of a foreground
population within $\sim 15$ kpc of the 
LMC (which is intriguingly close to the LMC's apparent tidal radius, 
$\gtsim 13$ kpc; Schommer \etal 1992)
that contributes significantly to the microlensing
optical depth is consistent with the current data. Other
populations (a thick LMC or material behind the LMC) will be even
more difficult to constrain. 
We close by noting that no available data yet compel us to 
adopt either an intervening stellar or halo MACHO interpretation
for all of the lenses. 
At least some lens events are 
asociated with the Clouds themselves (MACHO-98-SMC-1
with high certainty, Rhie \etal 1999; MACHO-LMC-9 with
less certainty, Bennett \etal 1996; Aubourg \etal 1999).
Similar direct identification of even a few true 
halo lenses would have significant implications
for the study of stellar evolution, galaxy formation, and
baryonic dark matter.

\vskip 1in
\noindent
ACKNOWLEDGMENTS: DZ acknowledges financial support from an
NSF grant (AST-9619576), a NASA 
LTSA grant (NAG-5-3501), a David and Lucile Packard Foundation
Fellowship and a Sloan Fellowship. DZ thanks Karl Gebhardt for the
use of his maximum likelihood dispersion measurement code and Ann
Zabludoff for comments on drafts and discussions. Lastly, we thank
David Alves, Jean-Philippe Beaulieu, Dave Bennett, and
Carme Gallart for being gracious and providing 
helpful comments even though they disagree with the basic premise
that the VRC stars are a significant lensing population.

\vskip 1cm
\noindent
{\centerline{\bf References}}

\apj{Alcock, C. \etal 1997a}{486}{697}

\apjlett{Alcock, C. \etal 1997b}{490}{59L}

\apjlett{Alcock, C. \etal 1998a}{499}{9L}

\refbook{Alcock, C. \etal 1998, ApJ, submitted (astro-ph/9807163)}

\refbook{Alfonso, C. \etal 1998 in press}

\aa{Ansari, R. \etal 1996}{314}{94}

\nature{Aubourg, E., \etal 1993}{365}{623}

\refbook{Aubourg, E., \etal 1999, AA, in press (astro-ph/9901372)}

\refbook{Beaulieu, J.-P., \& Sackett, P.D. 1998, in press}

\apjlett{Bennett, D.P. 1998}{493}{L79}

\refbook{Bennett, D.P. 1998, astro-ph/9808121}

\refbook{Berterlli, G., Bressan, A., Chiosi, C., Fagotto, F., \& Nasi,
E.
1994, A\&AS, 106, 275}

\refbook{Binney, J., \& Tremaine, S. 1987, galactic Dynamics
(Princeton:
Princeton Univ. Press)}

\aj{Bothun, G.D., \& Thompson, I.B. 1988}{96}{877}

\apjsup{Castellani, V., Chieffi., A., \& Pulone, L. 1991}{76}{911}

\apjlett{De Marchi, G., \& Paresce, F. 1997}{476}{L19}

\aj{Dohm-Palmer, R.C. et al. 1997}{114}{2527}

\aj{de Vaucouleurs, G. 1957}{62}{69}

\refbook{Efremov, Yu.N. 1978, AZ, 55, 272}

\refbook{Evans, N.W., Gyuk, G., Turner, M.S.,  \& Binney, J.J.
1997, Nature, in press}

\mnras{Feast, M.W., Thackeray, A.D., \& Wesseling, A.J., 1960}{120}{64}

\mnras{Feast, M.W., Thackeray, A.D., \& Wesseling, A.J., 1960}{121}{433}

\apj{Fitzpatrick, E.L. 1985}{299}{219}

\aa{Fusi Pecci, F., Ferraro, F.R., Crocker, D.A., Rood, R.T., \
Buonanno,
R. 1990}{238}{95}

\apj{Gallart, C. 1998}{495}{43}

\mnras{Gardiner, L.T., \& Noguchi, M. 1996}{278}{191}

\aj{Geha, M. \etal 1998}{115}{1045}

\refbook{Geisler, D., Bica, E., Dottori, H., Clari\'a, Piatti, A.E., 
\& Santos, J.F.C. 1998,
in IAU 190 the Magellanic Clouds (ed. Y.-H. Chu, N. Suntzeff,
J. Hesser, \&
D. Bohlender)}

\apj{Gould, A. 1998}{499}{728}

\refbook{Graff, D. and the EROS 2 Collaboration 1998,
in IAU 190 the Magellanic Clouds (ed. Y.-H. Chu, N. Suntzeff,
J. Hesser, \&
D. Bohlender)}

\refbook{Grebel, E.K. \& Brandner, W. 1998, in {\it The Magellanic
Clouds and Other Dwarf Galaxies}, eds. T Richtler \& J. Braan, in press}

\aj{Harris, J., Zaritsky, D., \& Thompson, I. 1997}{114}{1933}

\aj{Holtzman, J., \etal 1997}{113}{656}

\aj{Hughes, S.M.G., Wood, P.R., \& Reid, N. 1991}{101}{1304}

\refbook{Ibata, R.A., Lewis, G.F., \& Beaulieu, J.-P. 1998, ApJL, submitted}

\annrev{Iben, I., \& Renzini, A. 1983}{21}{271}

\apj{Johnston, K.V. 1998}{495}{297}

\apj{King, C.R., Da Costa, G.S., \& Demarque, P. 1985}{299}{674}

\aj{Kent, S.M. 1986}{91}{1301}

\refbook{Kim, S., Stavely-Smith, L., Dopita, M.A., Freeman, K.C.,
Sault, R.J., Kesteve, M.J., \& McConnell, D. \etal 1998, ApJ, in press}

\aa{Kippenhahn, R., \& Smith, L. 1969}{1}{142}

\apjlett{Kunkel, W.E., Demers, S., Irwin, M.J., \& Albert, L. 1997}
{488}{L129}

\aa{Luks, Th., \& Rohlfs, K. 1992}{263}{41}

\refbook{Majewski, S.R., Ostheimer, J.C., Kunkel, W.E., Johnston,
K.V., 
\& Patterson, R.J. 1998, in IAU 190 the Magellanic Clouds (ed. Y.-H. Chu, N. Suntzeff,
J. Hesser, \&
D. Bohlender)}

\apj{Massey, P., Lang, C.C., DeGioia-Eastwood, K., \& Garmany, C.D. 1995}{438}{188}

\apj{Meatheringham, S.J., Dopita, M.A., Ford, H.C., \& Webster,
B.L. 1988}{327}{651}

\apj{Paczy\'nski, B. 1986}{304}{1}

\refbook{Payne-Gaposchkin, C.H. 1971, Smithsonian Contr. Ap., No. 13}

\refbook{Putnam, M. \etal 1998, Nature, in press}

\aa{Renault, C. \etal 1997}{324}{69L}

\refbook{Rhie, S.H. \etal, 1999, astro-ph/9812252}

\pasp{Sahu, K.C. 1994}{106}{942}

\apj{Salpeter, E.E. 1955}{121}{161}

\refbook{Scalo, J.M. 1986, Fund. Cosmic Phys., 11, 1}

\aj{Schild, R.E. 1977}{82}{337}

\aj{Schommer, R.A., Olszewski, E.W., Suntzeff, N.B., and Harris, H.C.
1992}{103}{447}

\refbook{Shapley, H., 1956, American Scientist, 44, 73}

\refbook{Shectman, S.A., Schechter, P.L., Oemler, A.A., Tucker,
D., Kirshner, R.P., \& Lin H. 1992, in Clusters and Superclusters
of Galaxies, ed. A.C. Fabian (Dordrecht:Kluwer), 351}

\apj{Spitzer, L., \& Schwarschild, M. 1951}{114}{385}

\apj{Spitzer, L., \& Schwarschild, M. 1953}{118}{106}

\refbook{Stetson, P.B. 1997, Baltic Astron., 6, 3}

\refbook{Udlaski, A., Kubiak, M., \& Szymanski, M. 1997, Acta. A., 47, 319}

\apj{Villumsen, J.V. 1985}{290}{75}

\refbook{Vogt, N.P. 1995, Ph.D. Dissertation, Cornell University}

\aa{Wielen, R. 1977}{60}{263}

\aj{Zaritsky, D., Harris, J., \& Thompson, I. 1997}{114}{1002}

\aj{Zaritsky, D. \& Lin, D.N.C. 1997}{114}{2545}

\mnras{Zhao, H. 1998a}{294}{139}

\apjlett{Zhao, H. 1998b}{500}{149}

\vfill\eject\clearpage

.
\begin{figure}
\vskip -5in
\includegraphics{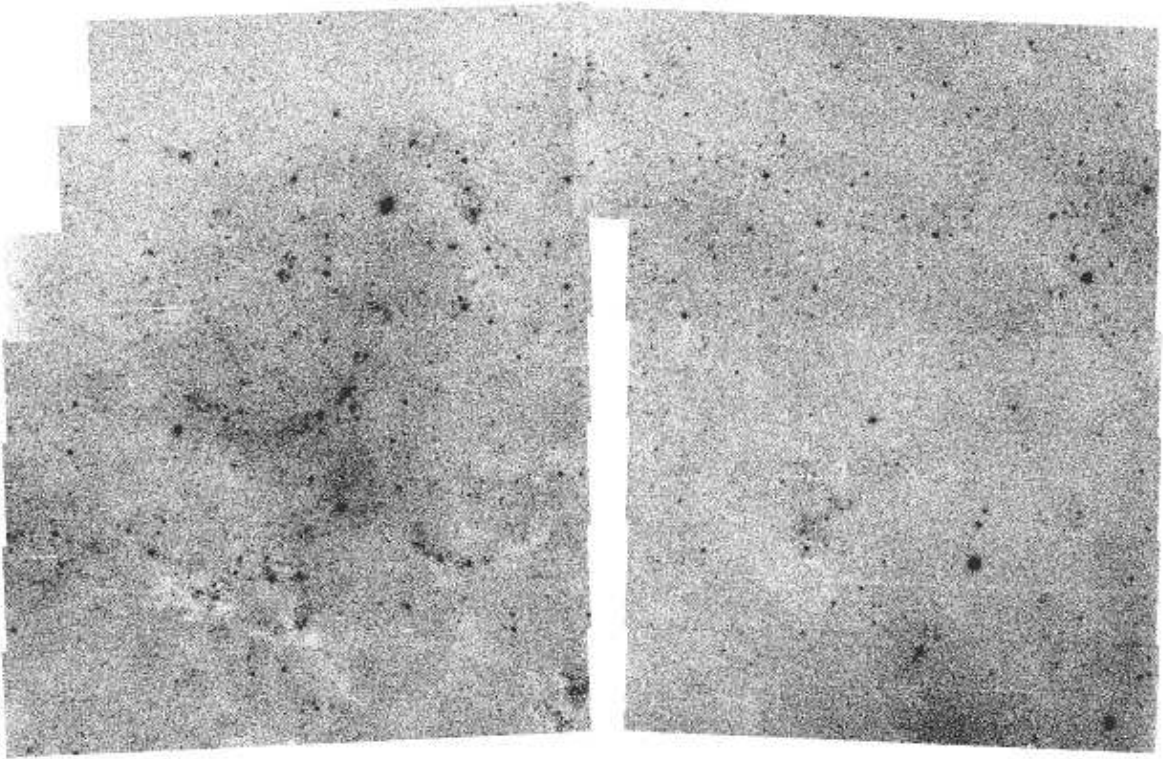}
\caption{
\noindent
The stellar density plot (for $V < 21$) for the region
in the LMC with reduced photometry from the Magellanic Cloud
Photometric Survey. The 
central coordinates are roughly $\alpha = 5^h20^m$ and 
$\delta = -66^\circ48^\prime$,
the image is $\sim 4^\circ$ wide, with North at the top and East
to the left. Each ``pixel'' corresponds to 30$^{\prime\prime}$.}
\end{figure}

\medskip
\clearpage\eject

\begin{figure}
\vskip 5in
\includegraphics{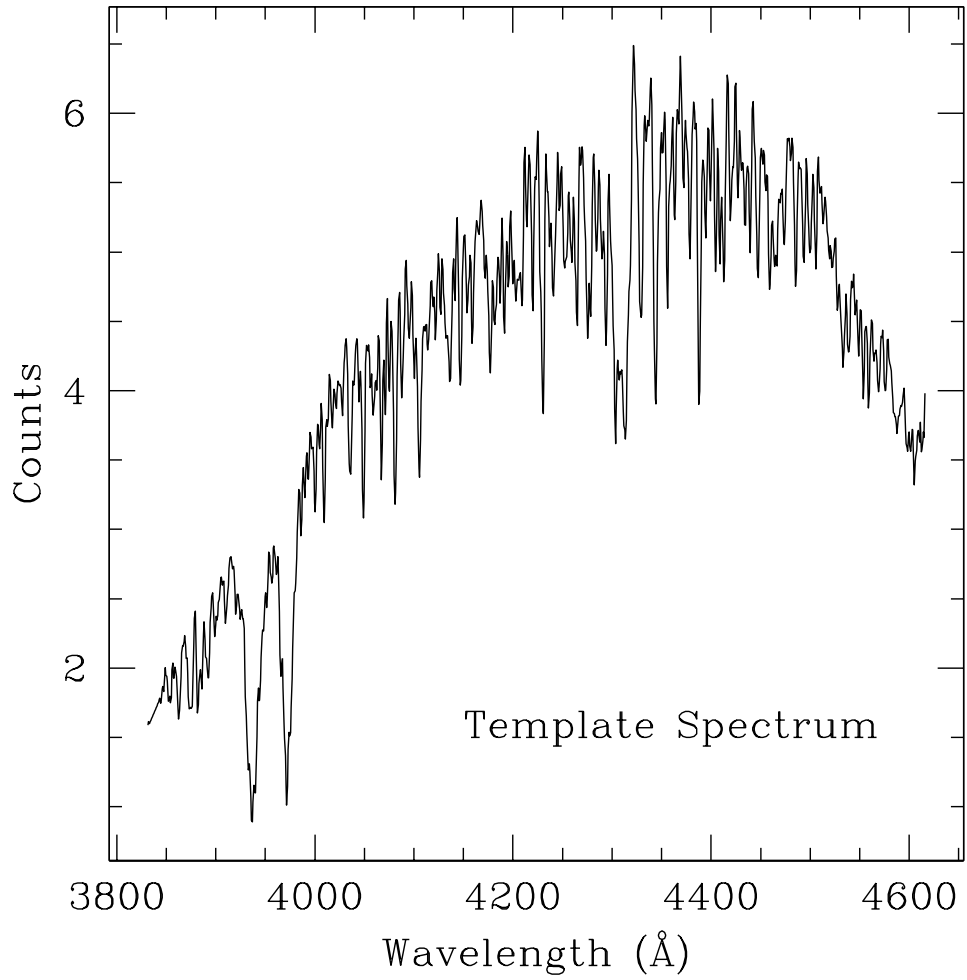}
\caption{
\noindent
The final velocity template spectrum (constructed from 
the sum of the VRC stellar spectra).}
\end{figure}
\medskip

\clearpage\eject
\begin{figure}
\vskip 5in
\includegraphics{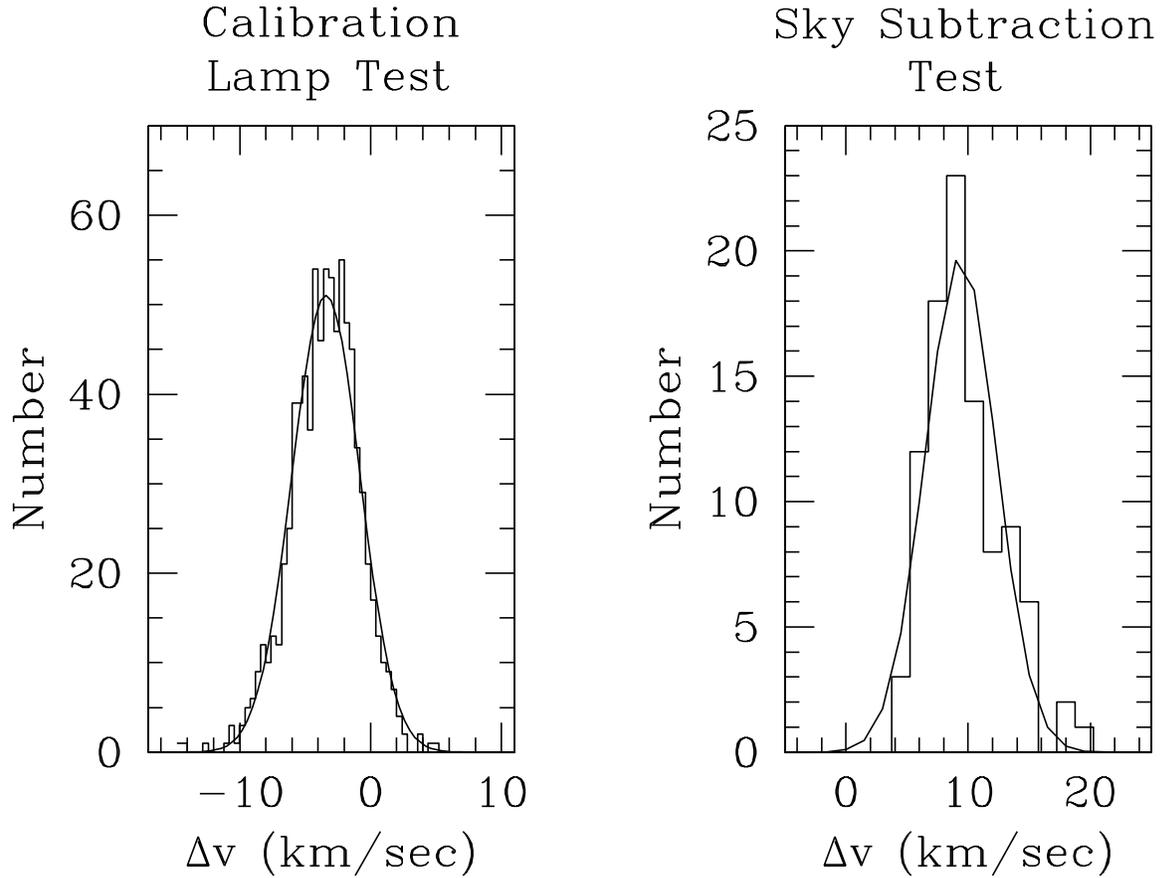}
\caption{
\noindent
Two tests of velocity uncertainties. The left panel shows
the results of measuring cross correlation velocities for one 
calibration spectrum against the others. The distribution is Gaussian
with low ($3$ km sec$^{-1}$) dispersion, demonstrating that the
wavelength solutions are stable. The right panel shows the difference
in the measured velocities with and without sky subtraction. 
The distribution is asymmetric but the dispersion of the fitted
Gaussian is only 5 km sec$^{-1}$. }
\end{figure}
\medskip

\clearpage\eject
\begin{figure}
\vskip 5in
\includegraphics{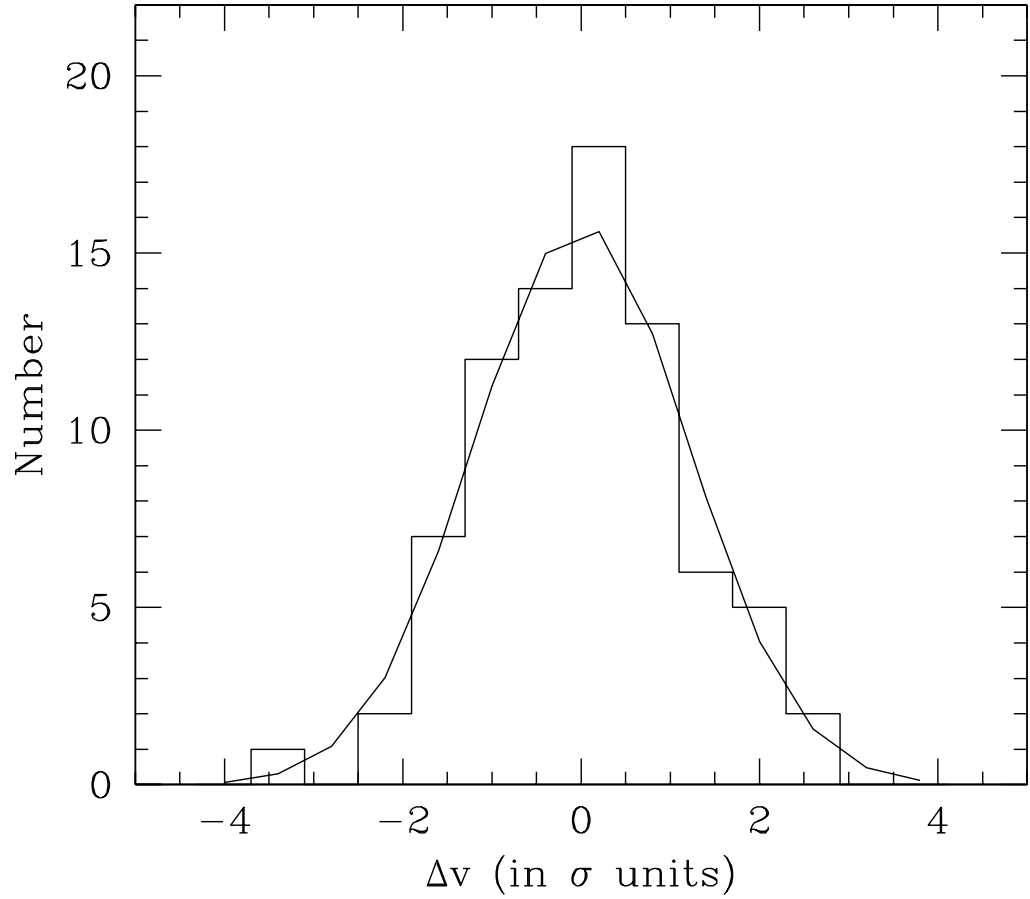}
\caption{
\noindent
Repeatability of velocities. We show the results of comparing
the radial velocity measured on one night vs. that measured on
another night, normalized by the uncertainty estimate produced by 
the cross-correlation software (IRAF XCSAO).}
\end{figure}
\medskip

\clearpage\eject
\begin{figure}
\vskip 5in
\includegraphics{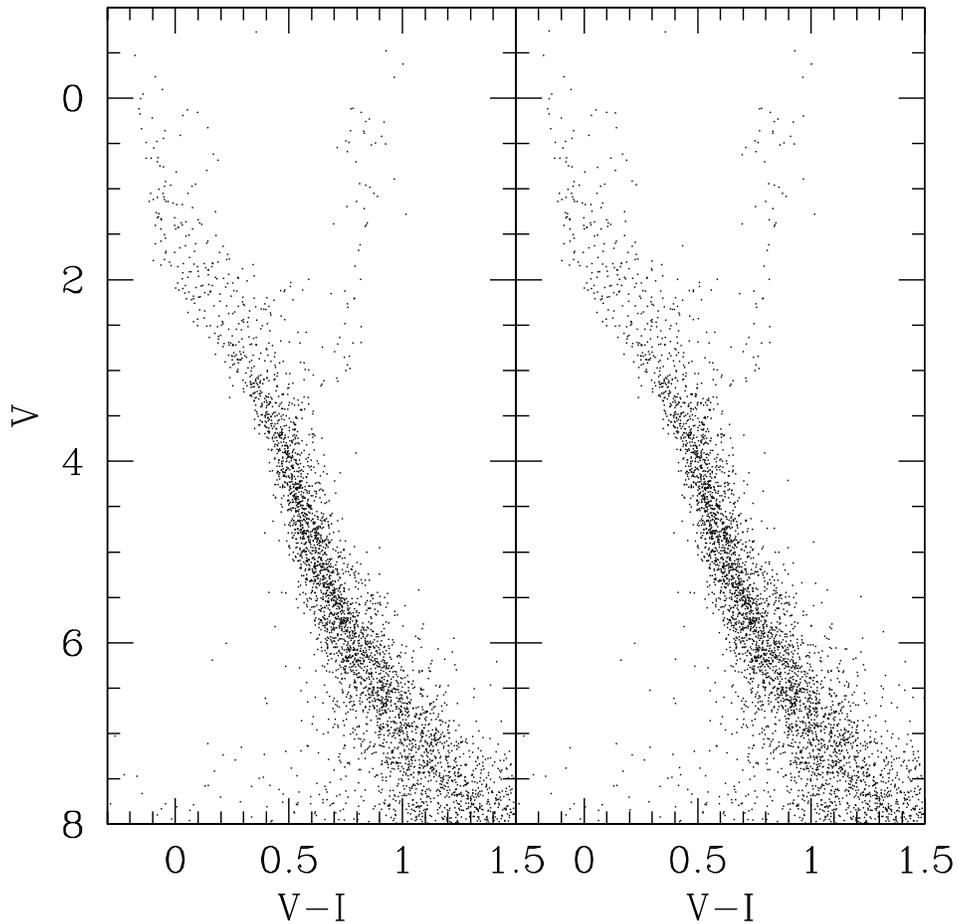}
\caption{
\noindent
The lower main sequences and a foreground population.
We compare the CMD obtained by Holtzman \etal (1997) in the left
panel to the same stellar population with the addition of a foreground
population (uniformly distributed at a distance between 35 and 40 kpc 
along the line of sight) that is 8\% of the LMC population.
The lack of obvious differences between the two CMDs
illustrates that even on the lower
main sequence, where the statistics are best, it is difficult to exclude
a foreground population.}
\end{figure}
\medskip

\clearpage\eject
\begin{figure}
\vskip 5in
\includegraphics{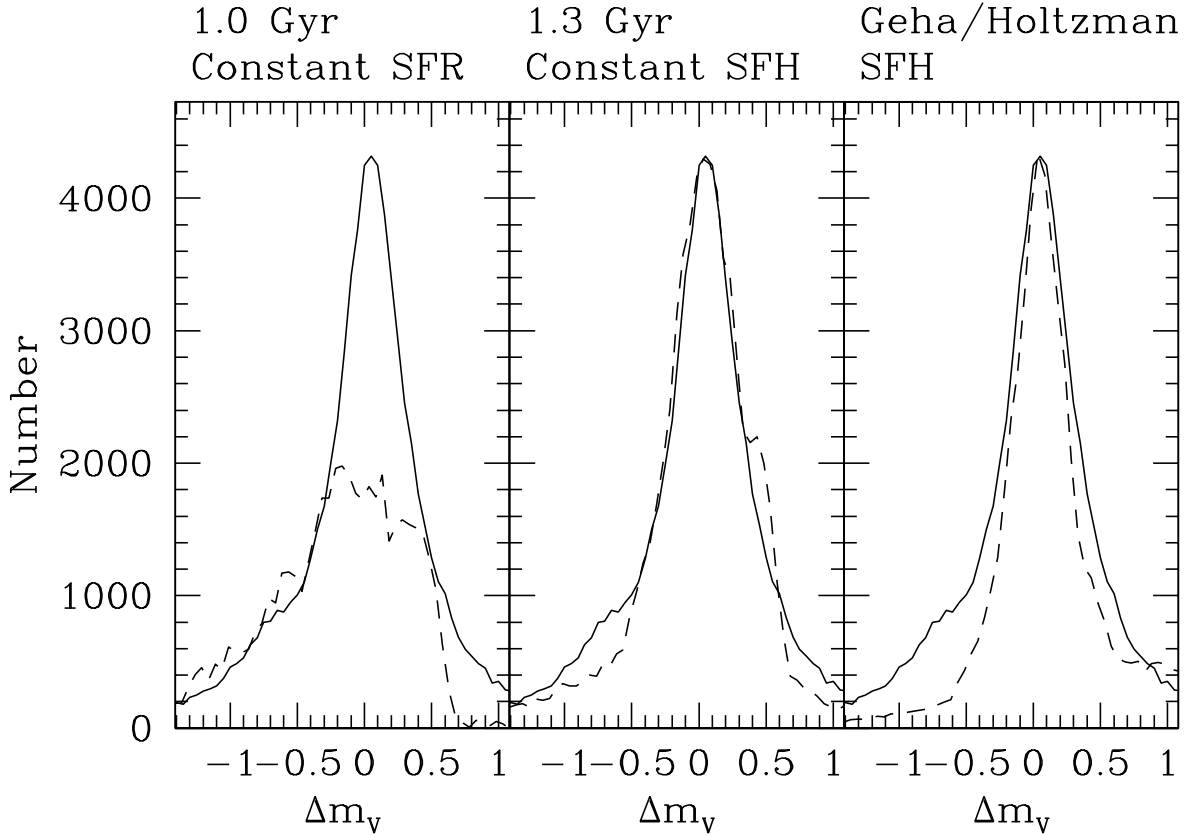}
\caption{
\noindent
The VRC/RC distribution. We plot histograms of vertical cuts
taken through CMDs at the color of the red clump. The wing toward the
left of the large peak is the VRC. The solid line are the data from 
ZL, the dashed lines represent results from three simulations. The
adopted star formation history is listed above each corresponding 
panel.}
\end{figure}
\medskip

\clearpage\eject
\begin{figure}
\vskip 5in
\includegraphics{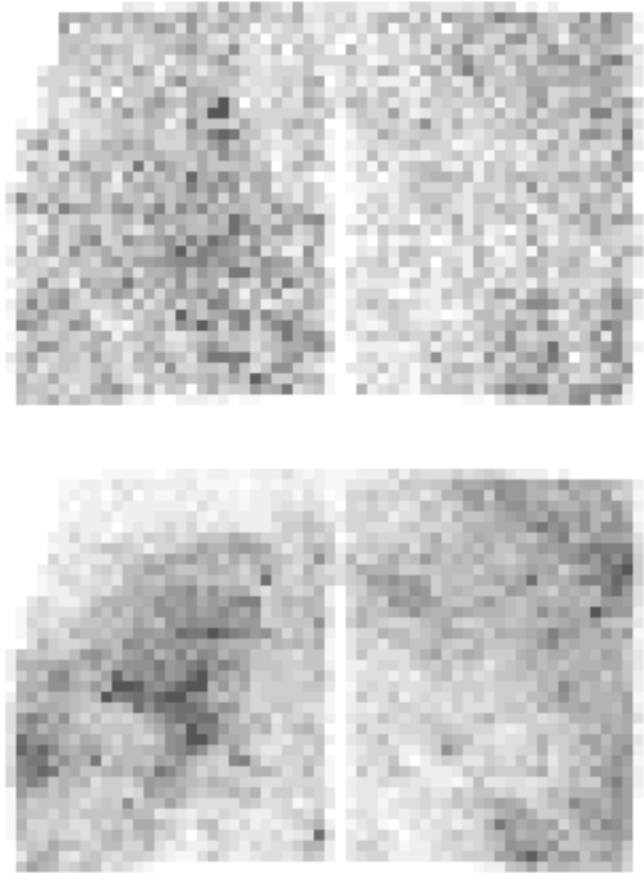}
\caption{
\noindent
The stellar density in $6^\prime \times 6^\prime$ pixels
for selected main sequence stars (19.15 $< V < 19.5$) in the lower
panel and the VRC in the upper panel. The total area shown in each
panel is the equivalent to that shown in Figure 1.}
\end{figure}
\medskip

\clearpage\eject
\begin{figure}
\vskip 5in
\includegraphics{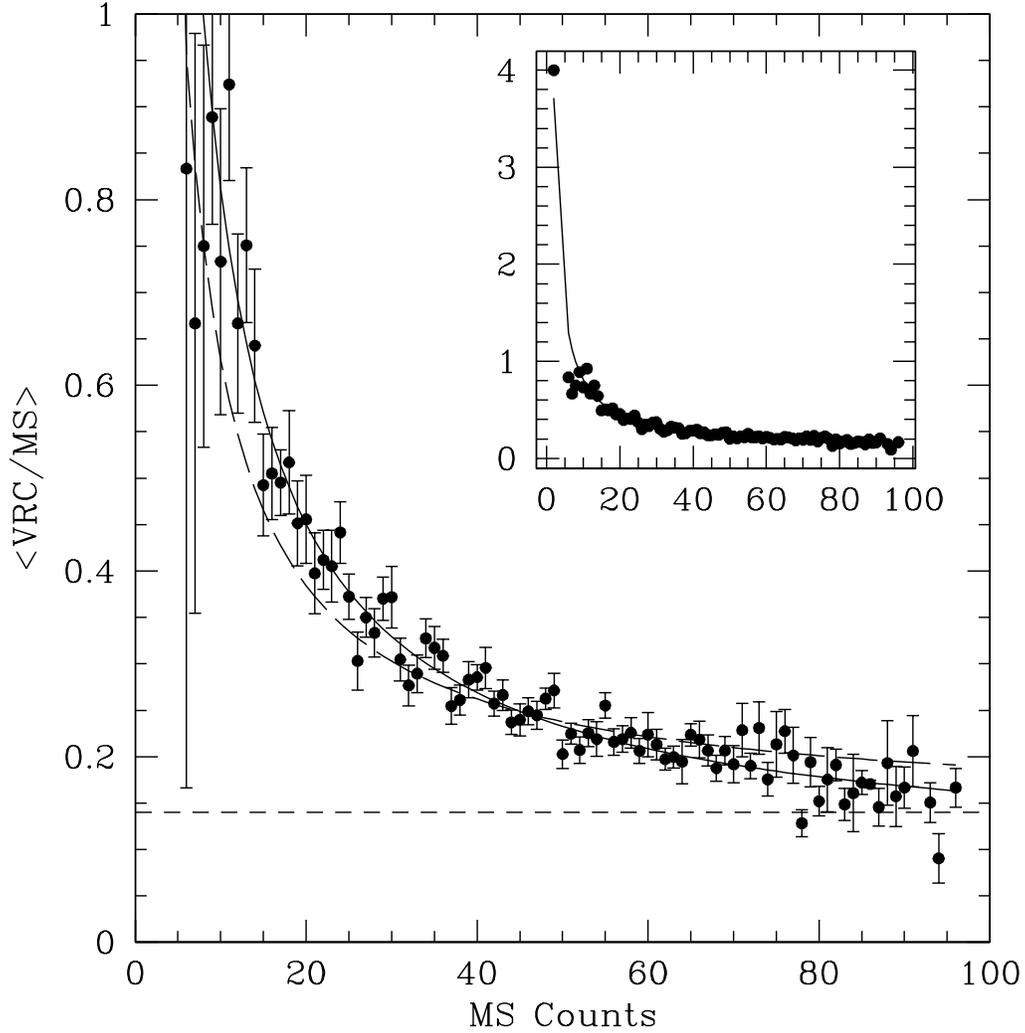}
\caption{
\noindent
The average ratio of the VRC stellar density to the
main sequence stellar density as defined in the text as a function
of main sequence stellar density 
(for $6^\prime \times 6^\prime$ boxes). 
The inset plot shows the entire
data, the principal plot shows more detail and includes all but one
of the data points. The solid curve represents the best fit model,
the long-dashed curve represents a model with a fixed contribution of
14\% of the MS stars to the VRC. The dashed horizontal line shows
the predicted asymptotic value of the ratio for the Geha \etal 
star formation history.}
\end{figure}
\medskip

\clearpage\eject
\begin{figure}
\vskip 5in
\includegraphics{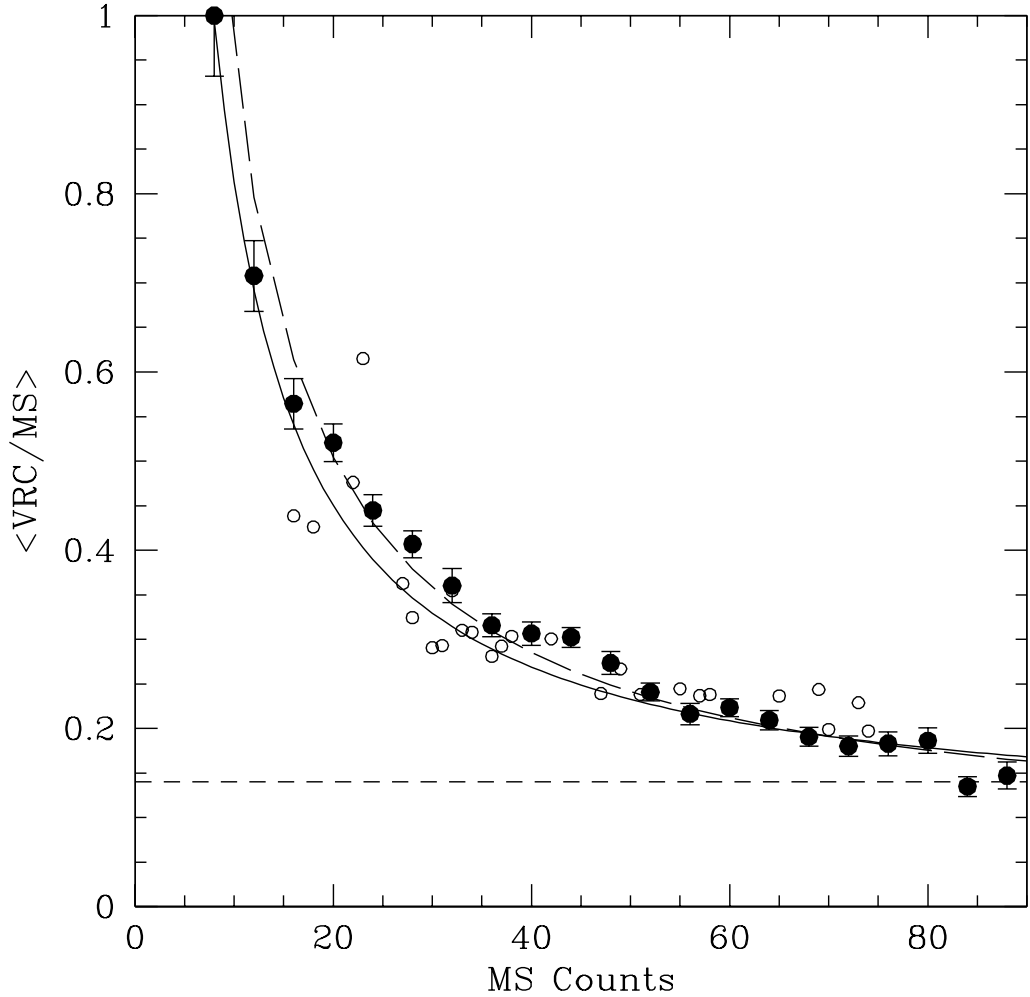}
\caption{
\noindent
The average ratio of the VRC stellar density to the
main sequence stellar density as defined in the text as a function
of main sequence stellar density (for $3^\prime \times 3^\prime$ boxes in
filled
circles and for $12^\prime \times 12^\prime$ boxes in open circles).
The number of MS stars have been normalized to represent the numbers found
per $6^\prime \times 6^\prime$ box.
The solid curve represents the best fit derived from the data in
Figure 8. 
The long-dashed curve represents the best fit model for the data from
$3^\prime \times 3^\prime$ boxes. The dashed horizontal line shows
the predicted asymptotic value of the ratio for the Geha \etal 
star formation history.}
\end{figure}
\medskip

\clearpage\eject
\begin{figure}
\vskip 5in
\includegraphics{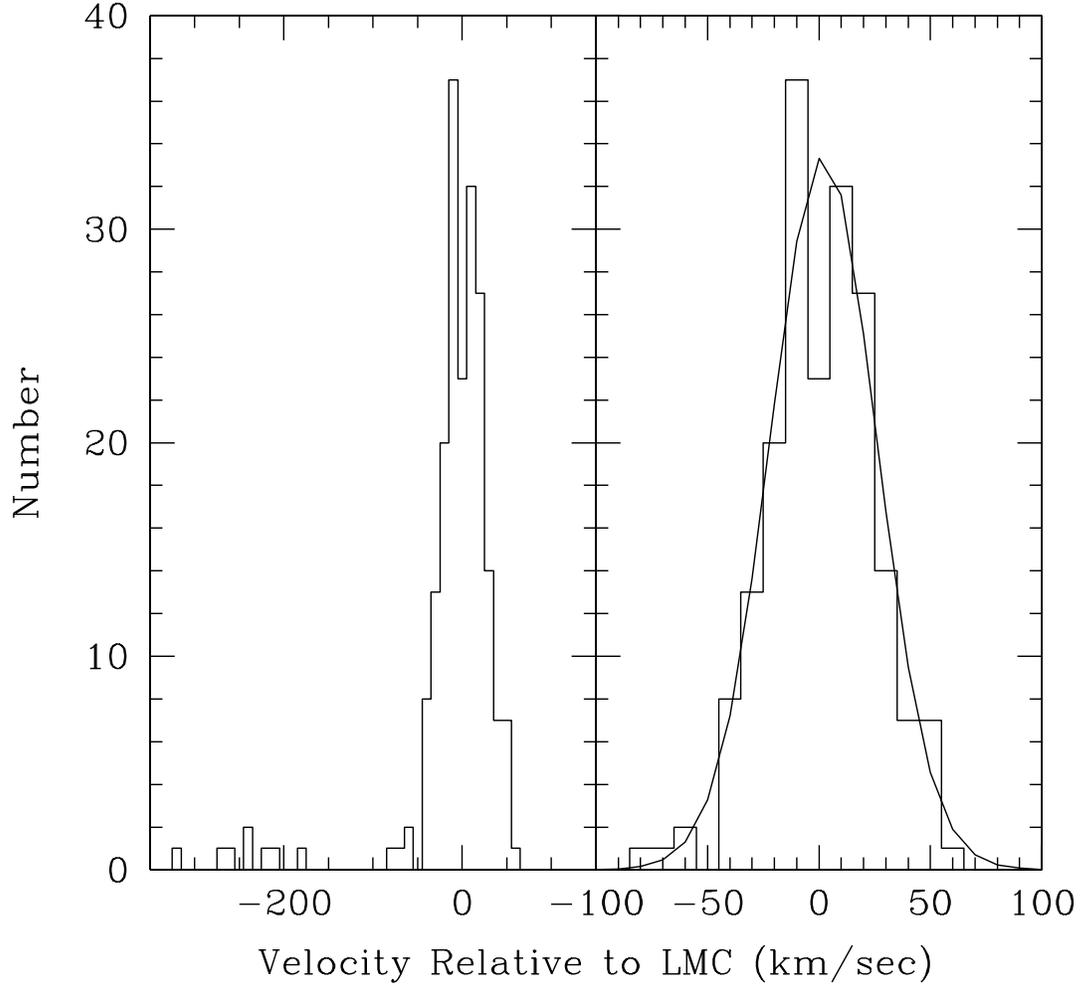}
\caption{
\noindent
The VRC radial velocity distributions.
In the left panel we show the radial velocities for all VRC candidates
with a significant correlation value. Due to the nature of the 
velocity template resolution, the velocity is with respect to the
mean of the large peak of VRC stars. In the right panel, we expand
around that peak and overplot the best fit Gaussian model.}
\end{figure}
\medskip

\clearpage\eject
\begin{figure}
\vskip 5in
\includegraphics{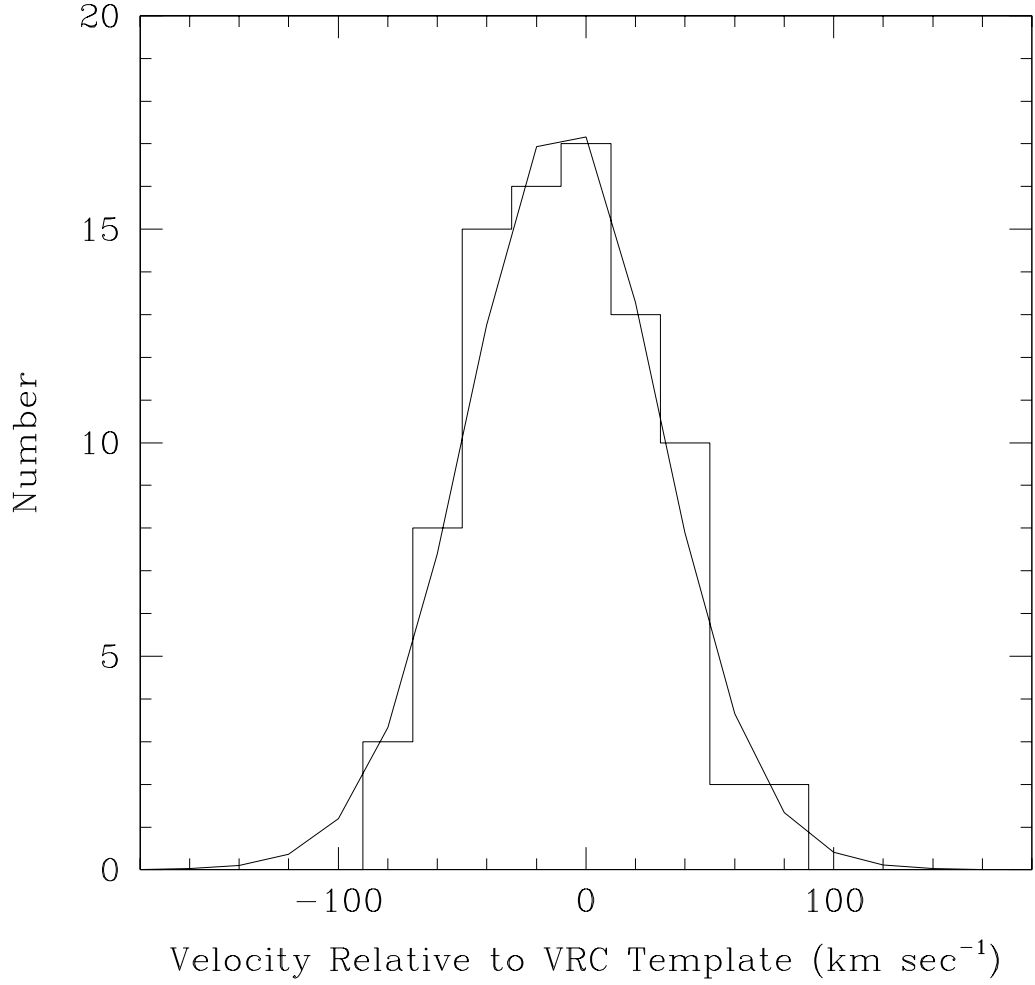}
\caption{
\noindent
The RC radial velocity distributions.
We show the radial velocities for all RC candidates
with a significant correlation value. Due to the nature of the 
velocity template resolution, the velocity is with respect to the
mean of the large peak of VRC stars (cf. Fig 10). The best fit Gaussian,
convolved with the observational uncertainties, is shown as the solid line.}
\end{figure}
\medskip

\clearpage\eject
\begin{figure}
\vskip 5in
\includegraphics{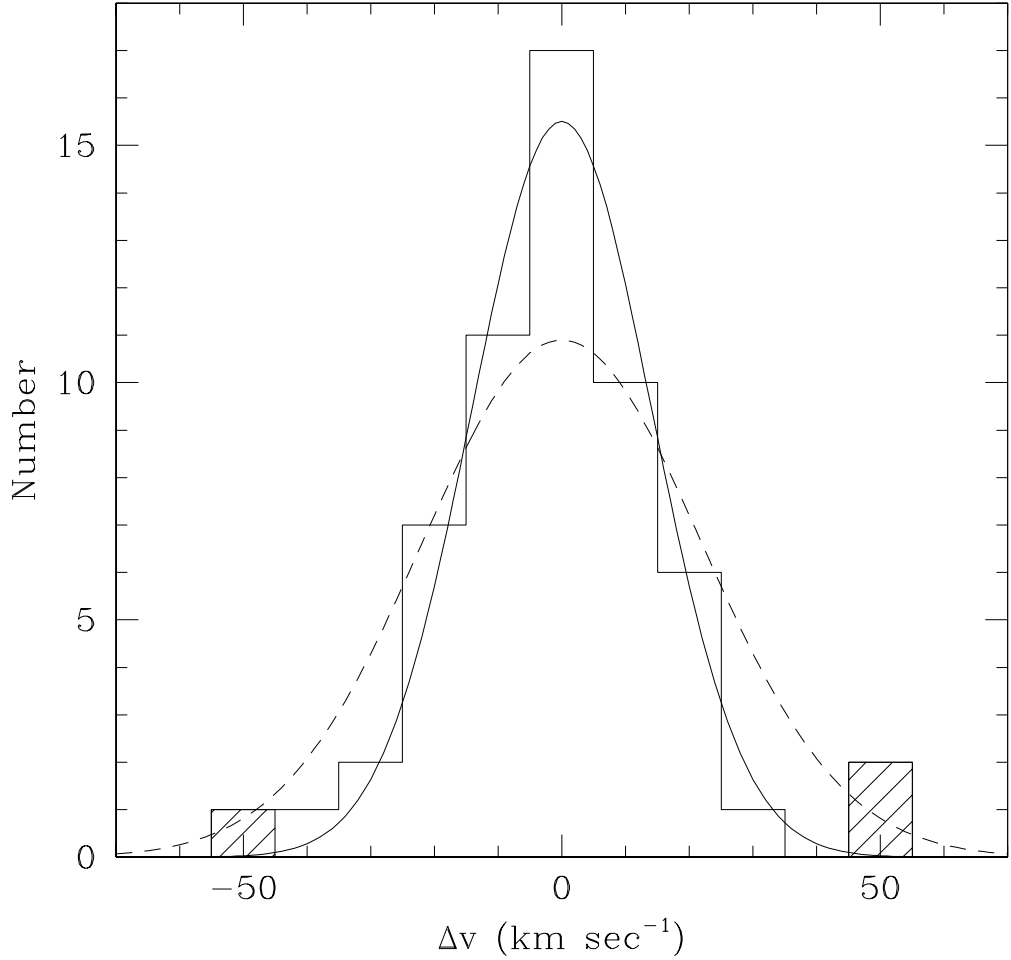}
\caption{
\noindent
The radial velocity distributions of supergiants
in the LMC (Feast, Thackeray, \& Wesselink 1960,1961) after
removing three stars within 1 kpc of the LMC center which had
$>$ 50 km sec$^{-1}$ residuals from the mean rotation curve (see text).
The dashed line shows the derived Gaussian for the entire population
of stars (intrinsic velocity dispersion = 19.6 km sec$^{-1}$). 
The shaded bins indicate highly deviant stars that are suspected
of distorting the derived Gaussian. The solid line shows the derived
Gaussian after removing the three stars in the shaded bins 
(intrinsic velocity dispersion = 10 km sec$^{-1}$).}
\end{figure}

\end{document}